\documentclass{emulateapj}
\usepackage{graphicx} 
\newcommand{\co}{$^{12}$CO}
\newcommand{\tco}{$^{13}$CO}
\newcommand{\jt}{${J=2-1}$}
\newcommand{\jo}{${J=1-0}$}
\newcommand{\kms}{km s$^{-1}$}
\newcommand{\msun}{${M_{\odot}}$}

\shorttitle{A CO line and Infrared Study of W51}
\shortauthors{Kang et al.}

\begin{document}

\title{A CO Line and Infrared Continuum Study of the Active Star-Forming Complex W51}
\slugcomment{Accepted for publication in ApJS}
\author{\sc Miju Kang\altaffilmark{1,2,3},
            John H. Bieging\altaffilmark{3},
            Craig A. Kulesa\altaffilmark{3},
            Youngung Lee\altaffilmark{1},
            Minho Choi\altaffilmark{1}, 
            and William L. Peters\altaffilmark{3}}
\altaffiltext{1}{International Center for Astrophysics,
                 Korea Astronomy and Space Science Institute,
                 838 Daedeokdaero, Yuseong, Daejeon 305-348, South Korea;
                 mjkang@kasi.re.kr}
\altaffiltext{2}{Department of Astronomy and Space Science,
                 Chungnam National University, Daejeon 305-764, South Korea}
\altaffiltext{3}{Steward Observatory,
                 University of Arizona,
                 933 North Cherry Avenue, Tucson, AZ 85721}

\begin{abstract}

We present the results of an extensive observational study of the active
star-forming complex W51 that was observed in the \jt\ transition of the
\co\ and \tco\ molecules over a 1\fdg25 $\times$ 1\fdg00 region with
the University of Arizona Heinrich Hertz Submillimeter Telescope. We
use a statistical equilibrium code to estimate physical properties of
the molecular gas. We compare the molecular cloud morphology with the
distribution of infrared (IR) and radio continuum sources, and find
associations between molecular clouds and young stellar objects (YSOs)
listed in {\it Spitzer} IR catalogs. The ratios of CO lines associated
with \ion{H}{2} regions are different from the ratios outside the
active star-forming regions. We present evidence of star formation
triggered by the expansion of the \ion{H}{2} regions and by cloud-cloud
collisions. We estimate that about 1\% of the cloud mass is currently
in YSOs. 
 
\end{abstract}



\section{Introduction}

Establishing the properties of molecular clouds is essential
in understanding the formation and evolution of star-forming
regions. Newly formed massive stars can affect the parental
molecular clouds through ionization, heating, and expansion of
the \ion{H}{2} regions, stellar winds, and supernova-driven shocks
\citep{Bieging09,Beuther07,Clark04,Gritschneder09,Povich09,Walborn02}.
These mechanisms are either compressing or dispersing the surrounding
clouds. For a better understanding of the feedback process in the
interstellar medium (ISM), detailed observations of the molecular clouds
and identification of embedded young stellar objects (YSOs) are required.

W51 is one of the most luminous star-forming regions in the first
quadrant of the Galactic plane. The high luminosity comes from a
large number of O-type stars \citep{Bieging75, Mehringer94} that
are within the molecular cloud \citep{Mufson79, Carpenter98}. In
the 21 cm radio continuum map \citep{Koo97}, W51 can be divided
into three regions: W51A (G49.4-0.3 and G49.5-0.4), W51B (G48.9-0.3,
G49.0-0.3, G49.1-0.4, G49.10-0.27, G49.17-0.21,and G49.2-0.4), and W51C
\citep{Kundu67,Bieging75,Mufson79,Koo97}. While W51C shows a nonthermal
continuum spectrum indicating a supernova remnant \citep{Subrahmanyan95},
W51A and W51B have thermal spectra indicating compact \ion{H}{2}
regions. The ionized gas in W51A is distributed in a broad velocity range
of 40 to 80 \kms\ \citep{Crampton78,Mehringer94,Pankonin79,Wilson70}. The
average systemic velocity of the ionized gas in W51B is $\sim$ 66
\kms\ \citep{Downes80,Pankonin79,Wilson70}. In the Massachusetts-Stony
Brook Galactic plane CO survey map covering the first Galactic quadrant
\citep{Sanders86} the W51 region is distinguished by bright CO emission
extending over a $1\arcdeg \times 1\arcdeg$ area centered on ($l, b$)
$\sim$ ($49\fdg5, -0\fdg2$) at the velocity range from 45 to 75 \kms\
\citep[see][Fig.1]{Carpenter98}. \cite{Carpenter98} studied this region
in detail in the \co\ and \tco\ \jo\ emission. They roughly divided the
molecular clouds associated with the \ion{H}{2} region complex into two
groups: the W51 giant molecular cloud (GMC) and the 68 \kms\ cloud. They
calculated a virial mass of $1.2\times10^6~M_\odot$ in the W51 GMC
and $1.9\times10^5~M_\odot$ in the 68 \kms\ cloud. Because the maximum
velocity permitted by the Galactic rotation \citep{Brand93} toward the
tangential point ($l=49\arcdeg$) of the Sagittarius spiral arm is $\sim
60$ \kms, the 68 \kms\ cloud is associated with a high-velocity (HV)
stream \citep{Burton70}. \cite{Carpenter98} speculated that the massive
star formation in W51 results from a collision between the W51 GMC and
the 68 \kms\ cloud which is related to a spiral density wave. \cite{Koo99}
and \cite{Kumar04} also argued that enhanced star formation in W51 might
be caused by the spiral density wave.

In this paper we present new observations of W51 in the \jt\
transition of \co\ and \tco\ lines. Our on-the-fly (OTF) maps are
fully sampled in two lines simultaneously with an angular resolution
of 32\arcsec, and with rms noise per velocity channel of $\sim$ 0.1 K
in antenna temperature. Our sensitive, fully sampled mapping allows
us to estimate mass more accurately and to delineate more detailed
structure of the molecular gas. The purpose of this study is to find
the correlation between the YSOs identified from the Galactic Legacy
Infrared Mid-Plane Survey Extraordinaire (GLIMPSE I) Catalogs and the
associated molecular clouds. We can approach the star formation history
in this massive star-forming complex by analyzing CO data. In Section
\ref{co:observations} we describe the observations and data processing. In
Section \ref{co:results} we present our analysis of the spectral features
and describe individual regions based on the CO, radio continuum, and
{\it Spitzer} data. In Section \ref{co:discussion} we discuss the spatial
distribution of the \ion{H}{2} regions and estimate the star formation
efficiency. In Section \ref{co:summary} we summarize our main conclusions.

\section{Observations and Processing}
\label{co:observations}

\subsection{\co\ and \tco\ \jt\ and \tco\ \jo}

The W51 GMC complex was mapped in the \jt\ transition of \co\ and \tco\
with the 10 meter Heinrich Hertz Telescope (HHT) on Mount Graham,
Arizona. The first observation was in 2005 November. A $1\fdg25 \times
1\fdg00$ mapping centered on ($l,b$)= ($49\fdg375, - 0\fdg200$) was
completed in 2008 February. The whole map consists of a total of 15
sub-fields. Each of the $15\arcmin \times 20\arcmin$ sub-fields was mapped
with OTF scanning in right ascension at 10\arcsec\ per second with
a row spacing of 10\arcsec\ in declination. Prior to 2006 December,
we used a single-polarization double sideband SIS mixer receiver with
system temperatures of typically 350 -- 500 K, depending on weather and
source elevation. Since 2006 December, the observations were made using
the 1.3 mm ALMA dual-polarization sideband-separating receiver with a
4-6 GHz intermediate frequency band. The receiver was tuned to the \co\
\jt\ line at 230.538 GHz in the upper sideband and the \tco\ \jt\ line
at 220.399 GHz in the lower sideband. Typical system temperatures were
in the range 180 -- 300 K. The spectrometers, one for each of the two
polarizations and the two sidebands, were filter banks with 256 channels
of 1 MHz width and separation (corresponding to a velocity resolution of
1.3 \kms\ for the \co\ \jt\ line). For some subfields of the \co\ \jt\
mapping, the spectra were measured with an acousto-optic spectrometer
(AOS) having a frequency resolution of 1 MHz, and a total bandwidth of
1 GHz. The main beam efficiency, measured from planets, was 0.85 with
an estimated uncertainty of ($+0.05$,$-0.10$).

The raw data were processed with the CLASS reduction package (from
the University of Grenoble Astrophysics Group). The intensity scales
for the two polarizations were determined from observations of W51D
made just before the OTF maps. System temperatures were checked by the
standard ambient temperature load method \citep{Kutner81} after every
other row of the map grid. Further analysis was done with the MIRIAD
software package \citep{Sault95}. Although the data observed early in the
program with a different receiver were noisier than the data obtained
using the ALMA receiver, data quality was improved after combining all
of the data using a ($1/T_{sys}$)$^{2}$ weighting. At the final step,
the data were convolved with a 16\arcsec\ (FWHM) Gaussian beam, giving a
36\arcsec\ effective resolution from the original angular resolution of
the telescope, 32\arcsec\ (FWHM). The two polarizations were averaged for
each sideband, yielding images with rms noise per velocity channel of 0.16
and 0.07 K in $T_A^*$ for the \co\ \jt\ and \tco\ \jt\ respectively. The
average rms noise level of the \co\ map is larger than \tco\ because
several sub-fields of \co\ map were observed just once with the noisier
receiver. The full set of HHT data will be made available in a separate
paper (Bieging et al. 2010, in preparation).

We obtained \tco\ \jo\ data for the same region from the Galactic Ring
Survey (GRS) \footnote{http://www.bu.edu/galacticring/} \citep{Jackson06}.
Their spectral resolution is 0.2 \kms, and the angular resolution and
sampling are 46\arcsec\ and  22\arcsec, respectively. We transform
antenna temperature of \tco\ \jo\ of GRS to main beam temperature,
dividing by a main beam efficiency of 0.48.

\subsection{{\it Spitzer} Data}

The Galactic Legacy Infrared Mid-Plane Survey Extraordinaire \cite[GLIMPSE
I;][]{Benjamin03} covered the Galactic plane ($ 10 \arcdeg <|l|<
65 \arcdeg,\, |b| < 1\arcdeg$) with the four mid-IR bands of the
Infrared Array Camera \cite[IRAC;][]{Fazio04} on the {\it Spitzer} Space
Telescope. Each IRAC band contains different spectral features : Band 1
(3.6 \micron) shows mainly continuum emission from stars, Band 2 (4.5
\micron) traces H$_{2}$ rotational transitions arising in the shocked
gas associated with outflows, and emission in Bands 3 (5.8 \micron) and
4 (8.0 \micron) is dominated by polycyclic aromatic hydrocarbon (PAH)
features. For this study, we have retrieved images of the 1\fdg25 $\times$
1\fdg00 region of W51 centered on ($l,b$) = ($49\fdg375, -0\fdg200$) by
combining two mosaic images which have a high resolution of 1\farcs2. For
the same region, we used the GLIMPSE I Catalog to extract point sources
that are detected at least twice in one band with an S/N $>$ 5. The
GLIMPSE I Catalog also tabulates $JHK_{s}$ flux densities from the Two
Micron All Sky Survey (2MASS) point source catalog \citep{Skrutskie06}
for all GLIMPSE sources with 2MASS identifications.

MIPSGAL \citep{Carey05} is a legacy program covering the inner Galactic
plane, $10\arcdeg < |l| < 65\arcdeg$ for $ |b| < 1\arcdeg$, at 24 and
70 \micron\ with the Multiband Imaging Photometer for {\it Spitzer}
\cite[MIPS;][]{Rieke04} Space Telescope. The resolution of the 24 \micron\
mosaics from the MIPSGAL survey is 2\farcs4. We extracted 24 \micron\
point-sources with $F/{\delta}F >$ 7, then bandmerged the 24 \micron\
sources with the GLIMPSE Catalog sources using a 2\farcs0 correlation
radius. The final source list contains data from all 8 bands: 2MASS
($JHK_s$), GLIMPSE (IRAC 1, 2, 3, 4 bands), and MIPSGAL 24 \micron. A
total of 104,582 sources within the target region were selected from the
GLIMPSE I Catalog. In an earlier paper, \cite{Kang09b} examined this set
of point sources to identify and classify YSOs near the W51 \ion{H}{2}
region complex using spectral energy distribution (SED) fits. All 8
bands are used for SED fitting as long as detected.

\section{Results}
\label{co:results}
\subsection{Velocity Structure}

Toward the W51 \ion{H}{2} region complex, most \co\ emission comes
from velocity intervals of $V_{LSR}=$ 0--25 \kms\ and 35--75 \kms\
\citep{Carpenter98}. We do not consider the former velocity component,
which originates from nearby molecular clouds. We focus on the later
velocity component, which is extended from 30 to 85 \kms\ in our sensitive
and small grid mapping, associated with the W51 \ion{H}{2} region complex.

\begin{figure}
\epsscale{1.2}
\plotone{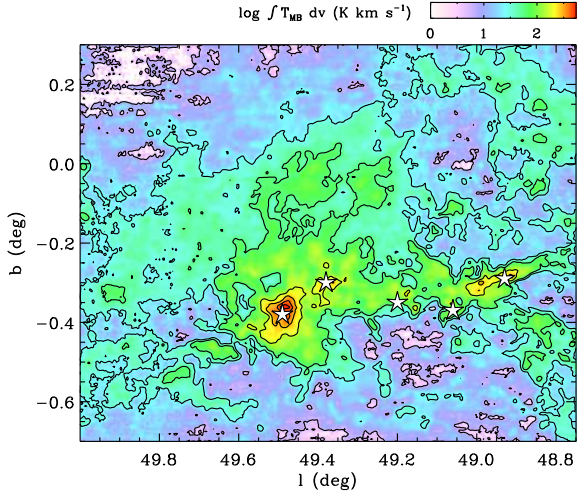}

\caption{Distribution of the \co\ \jt\ emission in the W51 \ion{H}{2}
region complex. The \co\ \jt\ emission is integrated from 30 to 85
\kms. The contour levels are 1, 5, 10, 30, 50, 70, and 90\% of the peak
value (542 K \kms) in the map. Star symbols mark the thermal radio
continuum sources listed by \cite{Goss70}. The names of the \ion{H}{2}
regions are G49.5-0.4, G49.4-0.3, G49.2-0.3, G49.1-0.4, and G48.9-0.3,
from east to west.}

\label{12co_lb}
\end{figure}

\begin{figure}
\epsscale{1.1}
\plotone{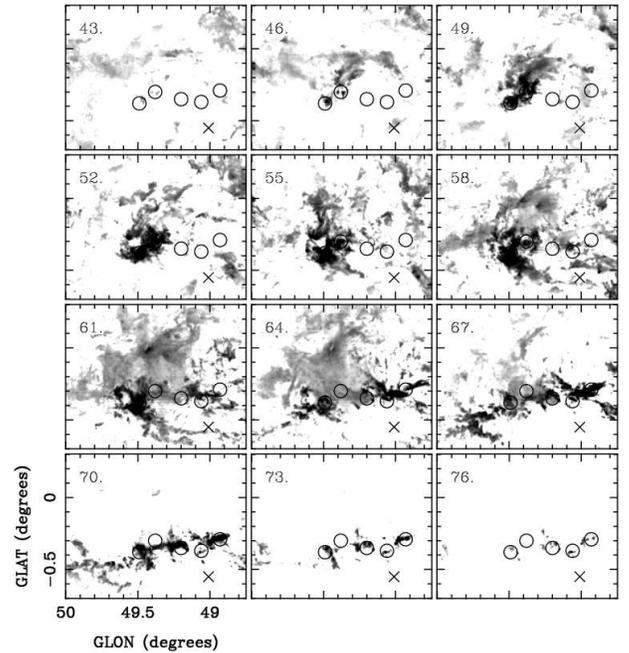}

\caption{Channel maps of the \co\ \jt\ emission averaged in 3 \kms\,
intervals over the velocity range from 43 to 76 \kms. The values printed
in the top left corner of each panel denote the center of the velocity
intervals. The gray scale in each panel is a logarithmic stretch from log
1.0 K (white) to log 10.0 K in $T_{MB}$ (black). Open circle symbols mark
the thermal radio sources, and cross symbol represents the non-thermal
radio source listed by \cite{Goss70}. }

\label{12cochanmap}
\end{figure}

\begin{figure}
\epsscale{1.1}
\plotone{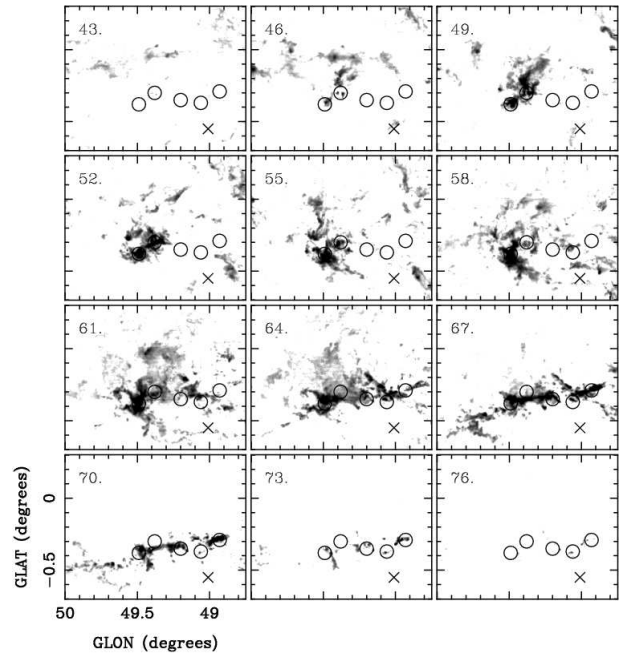}

\caption{Same as Fig. \ref{12cochanmap}, for \tco\ \jt. The gray scale
in each panel is a logarithmic stretch from log 0.3 K (white) to log
4.0 K in $T_{MB}$ (black). }

\label{13cochanmap}
\end{figure}

Figure \ref{12co_lb} represents the \co\ \jt\ intensity map integrated
over the velocity range between 30 and 85 \kms. Most of the bright
\co\ \jt\ emission peaks are coincident with bright radio continuum
sources. The W51 \ion{H}{2} region complex shows a complicated structure
in wide velocity ranges. \cite{Carpenter98} divided molecular clouds
associated with W51 into seven components at 44, 49, 53, 58, 60, 63,
and 68 \kms. In W51A, the 53 \kms\ component is associated with G49.4-0.3
and the 58, 60, 63 \kms\ components are associated G49.5-0.4. The 68 \kms\
cloud is related to W51B. Besides the bright emission associated with
radio continuum sources, there is \co\ \jt\ emission extended over the
whole of the $1\fdg25 \times 1\fdg00$ region. More detailed structures
of \co\ and \tco\ \jt\ emission are shown in channel maps (Figures
\ref{12cochanmap} and \ref{13cochanmap}).

In the 43 and 46 \kms\ channel maps of Figure \ref{12cochanmap}, there are
molecular clouds not associated with the radio continuum sources. Some
clouds are parallel to the Galactic plane at $b=0\arcdeg$, and the
others in the western\footnote{Note: In this paper, directions are with
reference to galactic coordinates, i.e., ``north'' means toward increasing
latitude and ``east'' means toward increasing longitude.} part of the map
are above the Galactic plane. CO emission in the W51A region appears in
the first channel, 43 \kms. Especially at the position of the brightest
radio continuum source, G49.5-0.4, the CO emission is distributed
through the whole velocity range from 46 to 76 \kms. In contrast, the
continuum sources of W51B are coincident with the filamentary structure
with velocity $\ge$ 64 \kms, which we discuss below. As pointed out by
\cite{Carpenter98}, the W51 molecular clouds with velocities of 58, 63
\kms\ are extended through a large area of the map, and the southern edge
of the diffuse emission seen in the 64 \kms\ channel map is seen against
the filamentary structure of 67 \kms. Two holes appear in the northern
($l,b = 49\fdg50, -0\fdg30$) and eastern ($l,b = 49\fdg60, -0\fdg35$)
parts of G49.5-0.4 from 49 to 64 \kms\ (Figure \ref{12cochanmap}).

\begin{figure}
\epsscale{1.2}
\plotone{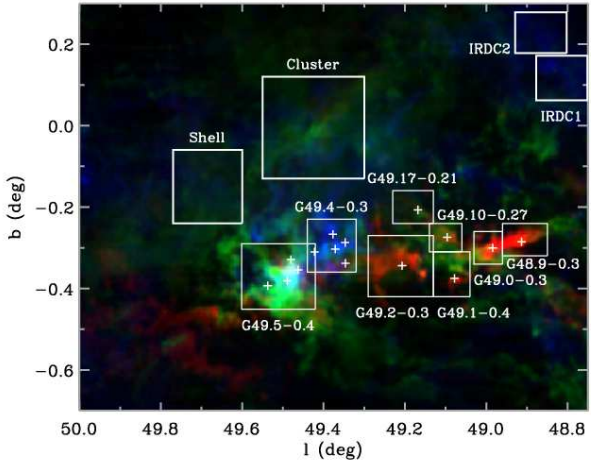}

\caption{Color image of the W51 region composed of \co\ \jt\, intensity
integrated over the velocity range of $30 - 55$ \kms\ (blue), $56 -
65$ \kms\ (green), and $66 - 85$ \kms\ (red). Cross symbols represent
the compact radio continuum sources listed by \cite{Koo97a}. White
boxes are the area for estimating the physical parameters in Table
\ref{CO.tbl.masses}.}

\label{W51_rgb_co}
\end{figure}

Figure \ref{W51_rgb_co} shows the total CO distribution around W51
in three velocity ranges, $30-55$, $56-65$, and $66-85$ \kms. The
molecular clouds associated with each radio continuum source show a range
of different velocity structure. The white boxes outline the regions
that we have analyzed to estimate the physical quantities from the \co\
and \tco\ \jt\ data.

\begin{deluxetable*}{lrrrrccccc}
\tabletypesize{\scriptsize}
\tablecaption{Physical parameters\tablenotemark{a}}
\tablewidth{0pt}
\tablehead{Name\tablenotemark{b} &
           \multicolumn{4}{c}{Selected area} &
           Velocity range\tablenotemark{c} &
           Mass\tablenotemark{d} &
           A$_{V}$\tablenotemark{e} &
           $^{13}R_{2-1/1-0}$\tablenotemark{f} &
           $^{13/12}R_{2-1}$\tablenotemark{g} \\
           &
           $l_\mathrm{min}$(\arcdeg) &
           $l_\mathrm{max}$(\arcdeg) &
           $b_\mathrm{min}$(\arcdeg) &
           $b_\mathrm{max}$(\arcdeg) &
           (\kms) &
           (10$^{4} $M$_\odot$) &
           (mag) &
           &
           }
\startdata
 G49.5-0.4  & 49.42&  49.60& $-$0.45  & $-$0.29& 49 -- 72 & 8.7 & 72 & 0.87 & 0.31\\
 G49.4-0.3  & 49.32&  49.44& $-$0.36  & $-$0.23& 47 -- 55 & 3.7 & 26 & 1.01 & 0.26\\
 G49.2-0.3  & 49.13&  49.29& $-$0.42  & $-$0.27& 62 -- 73 & 3.8 & 21 & 0.78 & 0.25\\
 G49.1-0.4  & 49.04&  49.13& $-$0.42  & $-$0.31& 56 -- 76 & 1.6 & 20 & 0.98 & 0.23\\
 G49.17-0.21& 49.13&  49.21& $-$0.24  & $-$0.18& 58 -- 69 & 0.9 & 20 & 0.78 & 0.37\\
 G49.10-0.27& 49.06&  49.14& $-$0.31  & $-$0.24& 58 -- 70 & 1.3 & 22 & 0.83 & 0.25\\
 G49.0-0.3  & 48.96&  49.03& $-$0.34  & $-$0.26& 62 -- 75 & 1.5 & 26 & 1.14 & 0.21\\
 G48.9-0.3  & 48.85&  48.96& $-$0.32  & $-$0.24& 63 -- 74 & 1.9 & 26 & 1.16 & 0.20\\
 Shell      & 49.60&  49.77& $-$0.24  & $-$0.06& 55 -- 68 & 2.9 & 15 & 0.43 & 0.12\\
 Cluster    & 49.30&  49.55& $-$0.13  &    0.12& 59 -- 69 & 6.0 & 15 & 0.41 & 0.13\\
 IRDC1      & 48.75&  48.88&    0.06  &    0.17& 47 -- 56 & 1.7 & 14 & 0.54 & 0.17\\
 IRDC2      & 48.80&  48.93&    0.18  &    0.28& 42 -- 49 & 1.1 &  9 & 0.52 & 0.17
\enddata

\tablenotetext{a}{ We assumed a distance of 6 kpc from the Sun to
derive the physical parameters.}
\tablenotetext{b}{ See Figure \ref{W51_rgb_co}}
\tablenotetext{c}{ Velocity range considered in the estimation of mass
and line ratios}
\tablenotetext{d}{ Mass derived using an LVG escape probability radiative transfer model}
\tablenotetext{e}{ Peak $A_{V}$ derived from the CO column density}
\tablenotetext{f}{ Mean \tco\ \jt\ /\ \jo\ intensity ratio}
\tablenotetext{g}{ Mean \tco\ /\ \co\ \jt\ intensity ratio}
\label{CO.tbl.masses}
\end{deluxetable*}

\subsection{Line Ratios}

\begin{figure*}
\plotone{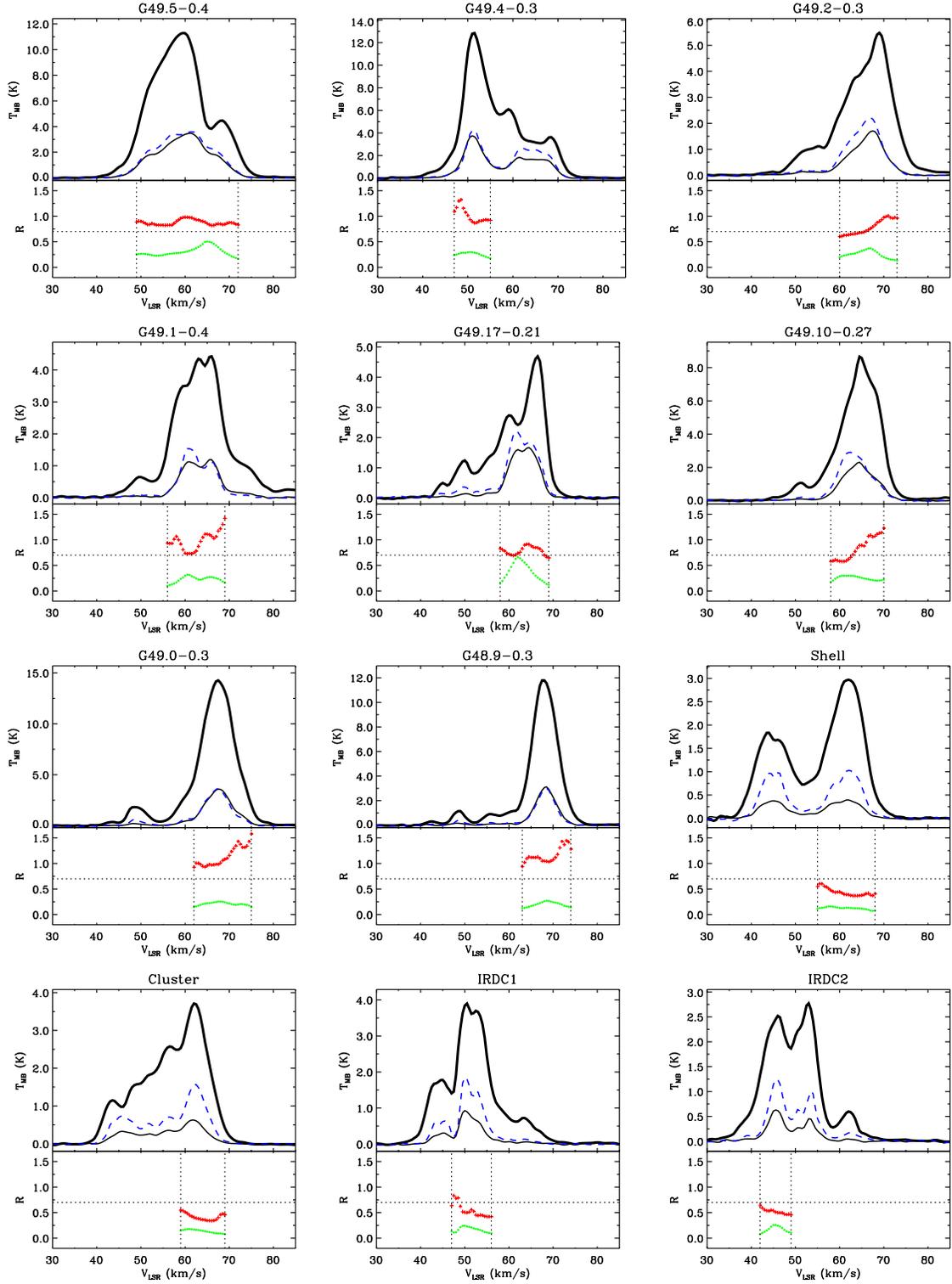}

\caption{Average spectra of \co\ \jt\ (thick line), \tco\ \jt\ (thin
line), and \tco\ \jo\ (dashed line) and ratios of $^{13}R_{2-1/1-0}$ (red
crosses) and $^{13/12}R_{2-1}$ (green dots) for the boxed areas in Figure
\ref{W51_rgb_co}. The horizontal dotted line is at 0.7 (see text). Two
vertical dotted lines show the velocity range for deriving mass and the
average line ratio of clouds, listed in Table \ref{CO.tbl.masses}. A
typical uncertainty is smaller than marker.}

\label{ave_spectra}
\end{figure*}

Figure \ref{ave_spectra} shows the average spectra and ratios for all
the regions marked with white boxes in Figure \ref{W51_rgb_co}. CO
line ratios are useful indicators of trends in line opacity and gas
column density (isotopologue ratios, e.g., \tco\ /\ \co\ \jt\ ($\equiv\
^{13/12}R_{2-1}$)); or of molecular excitation (e.g., \tco\ \jt\ /\ \jo\
($\equiv\ ^{13}R_{2-1/1-0}$)). As seen in Figure \ref{ave_spectra},
G49.5-0.4 has a wide velocity range, from 45 to 75 \kms\, showing a
peak at 60 \kms. The average $^{13}R_{2-1/1-0}$ value is 0.87 within
the two vertical dotted lines. The selected velocity ranges, covering
the strongest \tco\ emission, are associated with the \ion{H}{2}
regions or YSOs. The $^{13/12}R_{2-1}$ value increases with velocity
up to around 65 \kms\ and then decreases. The spectrum of G49.4-0.3
shows a peak at a lower velocity, 51 \kms. The spectrum in W51B has
strong emission between 55 and 75 \kms, which is different from the
molecular cloud associated with the W51A region. The $^{13}R_{2-1/1-0}$
values of G49.2-0.3, G49.1-0.4, G49.10-0.27, G49.0-0.3, and G48.9-0.3
in W51B systematically increase in the higher velocity ranges (Figure
\ref{ave_spectra}). The ratio plots in Figure \ref{ave_spectra}
generally show an anti-correlation between $^{13/12}R_{2-1}$ and
$^{13}R_{2-1/1-0}$. In Table \ref{CO.tbl.masses} we list the average
$^{13}R_{2-1/1-0}$ and $^{13/12}R_{2-1}$ values derived from each region.

The average $^{13}R_{2-1/1-0}$ ratios associated with all \ion{H}{2}
regions are greater than 0.7, which implies that these clouds are
relatively warm and dense. In contrast, the ratios of $^{13}R_{2-1/1-0}$
for the molecular clouds outside the active star-forming region, where
there are no known continuum sources, are smaller than 0.7 (see Shell,
Cluster, IRDC1, and IRDC2 in Figure \ref{ave_spectra}). Molecular
clouds with low ratios were often observed in dark clouds and outer
envelopes of giant molecular clouds \citep{Sakamoto97}.

\begin{figure}
\epsscale{1.1}
\plotone{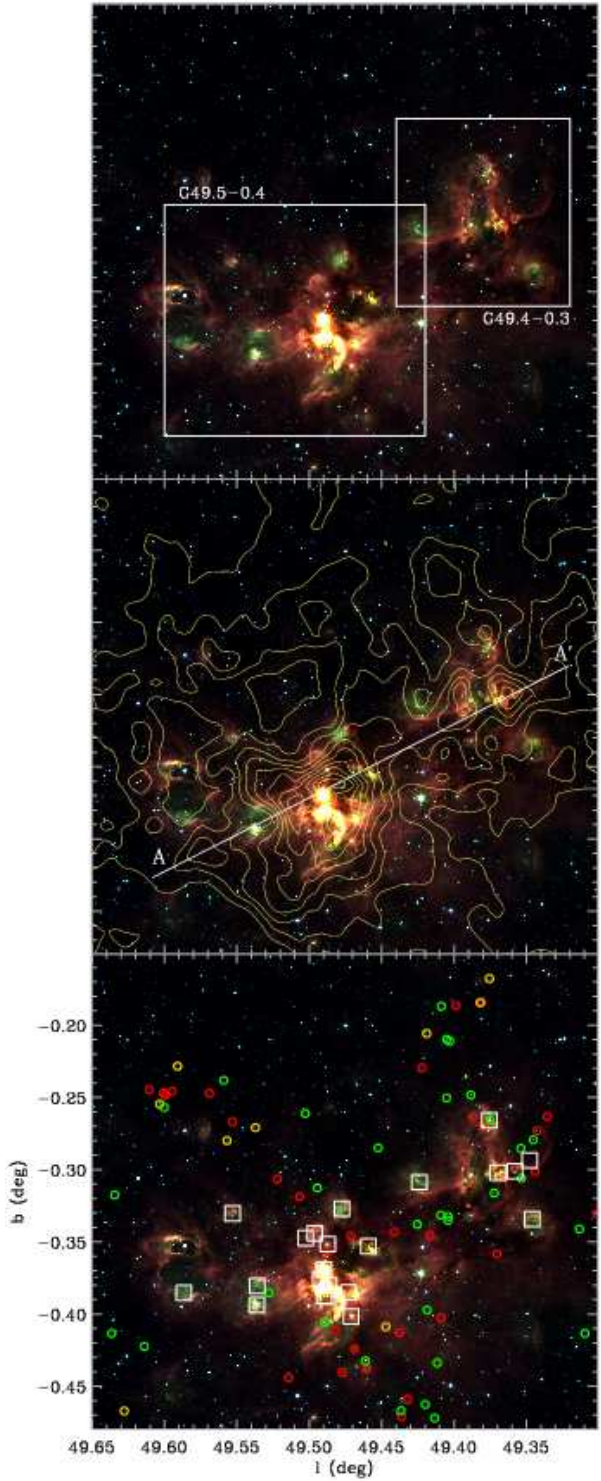}

\caption{{\it Top}: Color image of the W51A region composed of IRAC 5.8
\micron\ (red), 4.5 \micron\ (green), and 3.6 \micron\ (blue). White
boxes show the area for analyzing the labeled radio continuum source in
detail. {\it Middle}: Contour map of \co\ \jt\ intensity integrated from
45 to 65 \kms. The contour levels are 40, 80, 120, 160, 200, 240, 280,
320, 360, and 400 K \kms. The oblique line shows the cut for the PV maps
presented in Figure \ref{W51A_pvmap}. {\it Bottom}: Squares are compact
radio continuum sources listed by \cite{Mehringer94}. YSO candidates
(open circles) are marked in red for Stage 0/I, yellow for Stage II,
and green for ambiguous sources.}

\label{W51A_rgb}
\end{figure}

\begin{figure}
\epsscale{1.0}
\plotone{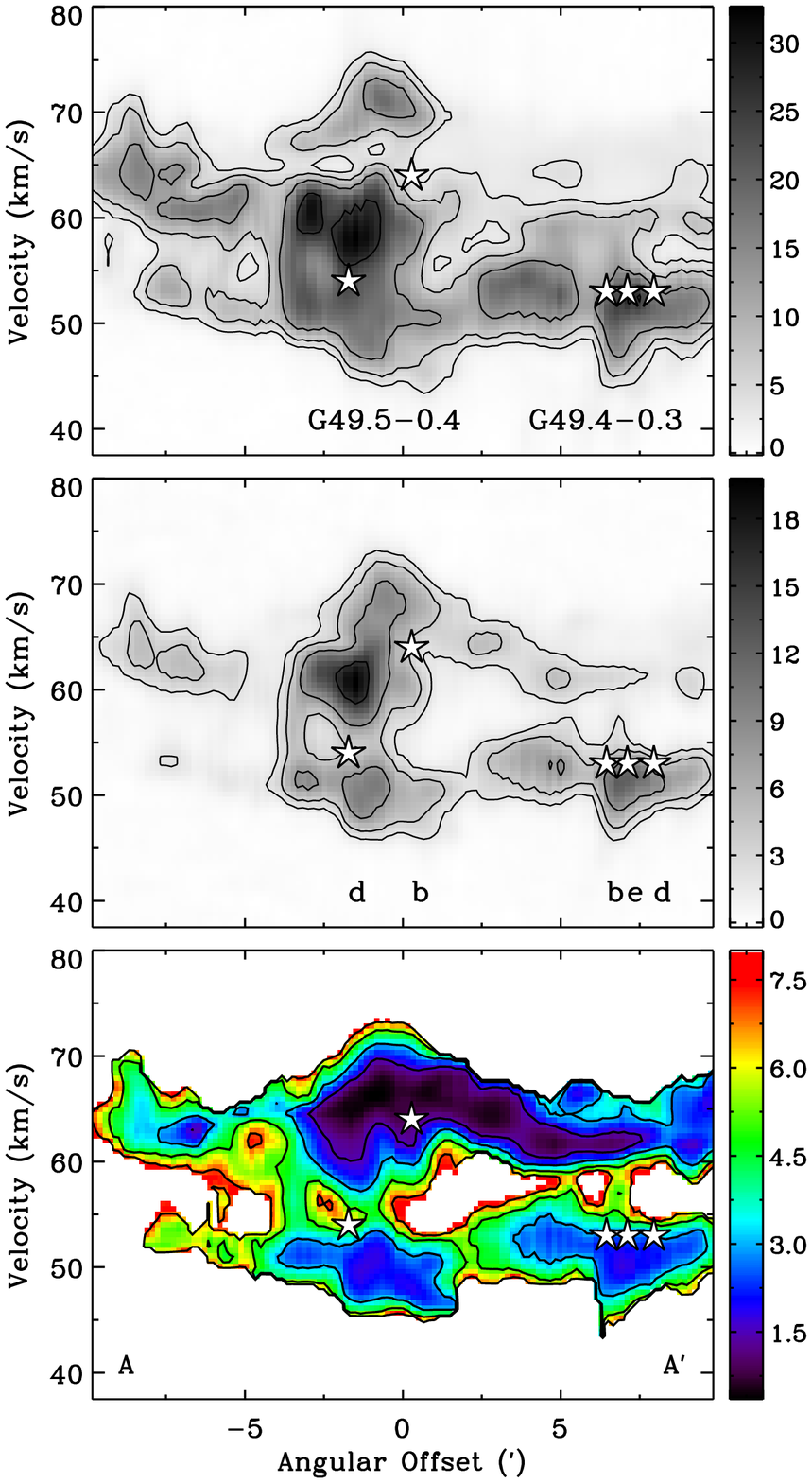}

\caption{\co\ \jt\ ({\it top}), \tco\ \jt\ ({\it middle}) and
$^{12/13}R_{2-1}$ ({\it bottom}) position-velocity map through the W51
A region from southeast (A) to northwest (A$^\prime$) marked at the
second panel of Figure \ref{W51A_rgb}. The temperature scale is indicated
on the right (K in $T_{MB}$). Contour levels of \co\ \jt\ and \tco\ \jt\
maps are 10, 20, 40, and 80\% of peak value in the map. Contour levels
of $^{12/13}R_{2-1}$ map are 1.5, 3.0, 5.0, and 7.0. Star symbols
represent the position of the radio continuum sources (d and b of
G49.5-0.4 and b, e, and d of G49.4-0.3). The velocities of the radio
continuum sources come from the radio recombination line observations
\citep{vanGorkom80,Pankonin79}.\\}

\label{W51A_pvmap}
\end{figure}

\subsection{Physical Properties}
\label{sec:physical properties}

We use a non-LTE statistical equilibrium treatment of the CO molecular
excitation to study physical properties of the molecular cloud. We apply
the escape probability radiative transfer and photodissociation model
of \cite{Kulesa05} to our \co\ and \tco\ \jt\ data. We calculate grids
of CO level populations for wide ranges of volume densities ($10^2 -
10^7$ cm$^{-3}$) and temperatures ($5 - 300$ K) assuming detailed
balance and steady state. From these model grids, the total CO column
density at each observed pixel is computed from the peak temperature,
line widths and integrated intensities of the observed CO lines. The
velocity range for integrating CO lines was determined from the range
covering the strongest \tco\ \jt\ emission on the line of sight (Figure
\ref{ave_spectra}). Assuming that the CO heating is dominated by photon
processes (e.g., the photoelectric heating of dust), a coarse estimate
of the incident radiation field can be made for each point in the
map. The photodissociation model is then applied to estimate a total
hydrogen column density from the CO data. This calculation is based on
the CO and H$_2$ photodissociation treatments of \cite{vanDishoeck88} and
\cite{Black87}, respectively, using a total interstellar carbon abundance
of $C/H=2.4\times10^{-4}$ \citep{Cardelli96}. Because the CO abundance
is a strong function of column density, blind application of a uniform CO
``dark cloud'' abundance ($10^{-4}$) can lead to a gross underestimate of
total hydrogen column density and gas mass. The visual extinction, $A_V$,
is calculated from $N(H)$ using $N(H)/A_V = 1.9 \times 10^{21}$ cm$^{-2}$
mag$^{-1}$ \citep{Bohlin78}. The distance to W51 derived in previous
studies vary from 2.0 to 8.3 kpc. \cite{Figueredo08} derived 2.0$\pm$0.3
kpc by spectroscopic parallaxes of O-type stars. The kinematic distance
is 5.5 kpc using radio recombination lines \citep{Russeil03}. The values
derived from water maser proper motion measurements are 6.1$\pm$1.3
kpc \citep{Imai02}, 7$\pm$1.5 kpc \citep{Genzel81}, and 8.3$\pm$2.5
kpc \citep{Schneps81}.  Recently, \cite{Xu09} found the distance of
$5.1^{+2.9}_{-1.4}$ kpc using trigonometric parallax measurements. In this
paper, we adopt a distance of 6 kpc and use that value to calculate the
gas masses of the selected regions shown in Figure \ref{W51_rgb_co}. In
Table \ref{CO.tbl.masses}, we list the size of the regions, velocity
range, masses, extinctions, and the mean ratios of the lines.

\begin{figure}
\epsscale{1.1}
\plotone{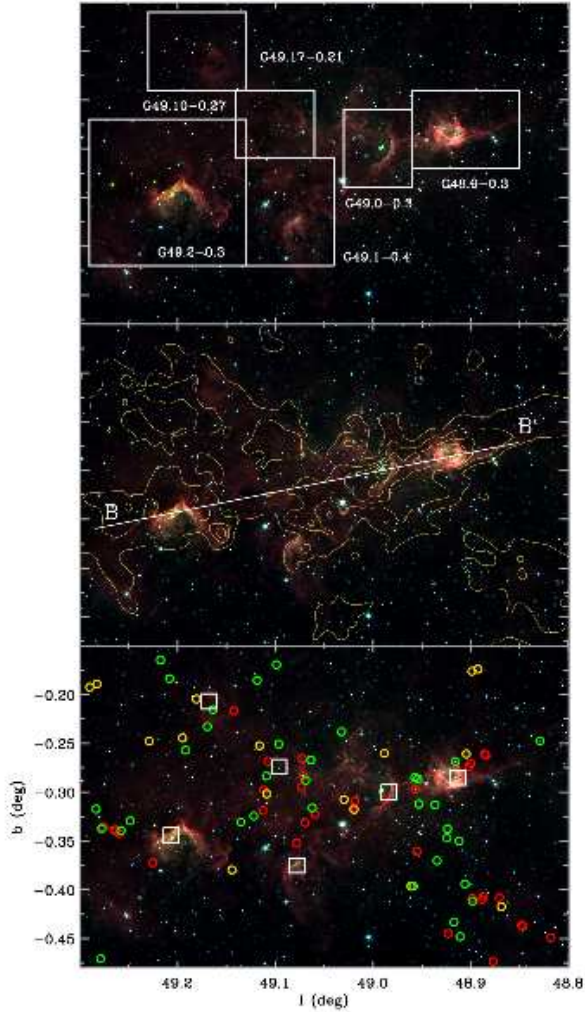}

\caption{{\it Top}: Color image of the W51B region composed of IRAC 5.8
\micron\ (red), 4.5 \micron\ (green), and 3.6 \micron\ (blue). White
boxes show the area for analyzing the labeled radio continuum source
in detail. {\it Middle}: Contour map of \co\ \jt\ intensity integrated
from 55 to 75 \kms. The contour levels are 30, 90, 150, 210, and 270
K \kms. The oblique line shows the cut for the position-velocity maps
presented in Figure \ref{W51B_pv}. {\it Bottom}: Squares are compact
radio continuum sources listed by \cite{Koo97a}. YSO candidates (open
circles) are marked in red for Stage 0/I, yellow for Stage II, and green
for ambiguous sources.}

\label{W51B_rgb}
\end{figure}

\begin{figure}
\epsscale{1.2}
\plotone{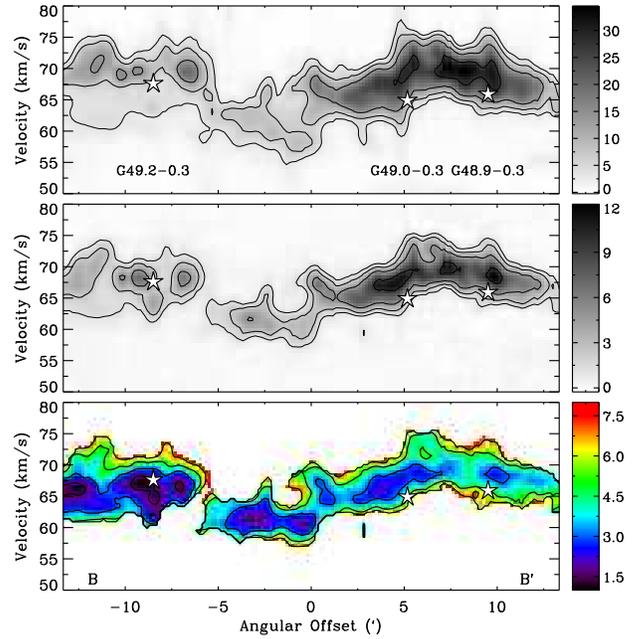}

\caption{\co\ \jt\ ({\it top}), \tco\ \jt\ ({\it middle}) and
$^{12/13}R_{2-1}$ ({\it bottom}) position-velocity map through the radio
continuum sources G49.2-0.3, G49.0-0.3, and G48.9-0.3 from B to B$^\prime$
in Figure \ref{W51B_rgb}. The temperature scale is indicated on the right
(K in $T_{MB}$). Contour levels of the \co\ \jt\ and \tco\ \jt\ maps are
10, 20, 40, and 80\% of the peak value in the map. Contour levels of the
$^{12/13}R_{2-1}$ map are 1.5, 3.0, 5.0, and 7.0. Star symbols represent
the radio continuum sources. The RRL velocities of radio continuum
sources are the mean velocities of the ionized gas \citep{Pankonin79}.}

\label{W51B_pv}
\end{figure}

\subsection{Large Scale Structure}
\subsubsection{W51A}

W51A includes the radio continuum sources G49.5-0.4 and G49.4-0.3 (Figure
\ref{W51A_rgb}). G49.5-0.4 is the brightest component in radio continuum
in the W51 region. It contains the infrared sources W51 IRS1 and IRS2
\citep{Wynn-Williams74}, and the H$_{2}$O masers W51 MAIN, W51N, and W51S
\citep{Genzel77}. \cite{Mehringer94} found that G49.5-0.4 consists of 16
discrete components using the VLA radio continuum observations at 3.6,
6, and 20 cm. G49.4-0.3 located to the west of G49.5-0.4 is associated
with 6 radio continuum sources \citep{Mehringer94}.

Figure \ref{W51A_rgb} shows the distribution of mid-infrared
and CO emission in W51A. The radio continuum sources detected by
\cite{Mehringer94} are marked with squares and YSO candidates are marked
with colored open circles. YSO candidates were identified and classified
by \cite{Kang09b} using the SED fitting tools of \cite{Robitaille07}. Each
color represents the evolutionary stage of the YSO candidates: red for
Stage I/0, yellow for Stage II, and green for ambiguous sources. Most YSO
candidates in the W51A region are Stage I/0 (red) or ambiguous sources
(green). The bright IRAC 4.5 \micron\ emission (green) is seen at the
position of the radio continuum sources. This 4.5 \micron\ emission
might be explained as line emission from H$_2$ in shocked molecular gas
\citep{Cyganowski08,Davis07,Shepherd07,Smith06}.

Figure \ref{W51A_pvmap} shows \co\ \jt, \tco\ \jt, and the ratio of
\co\ \jt\ to \tco\ \jt\ ($\equiv\ ^{12/13}R_{2-1}$) in the form of
position-velocity (PV) maps of G49.5-0.4 and G49.4-0.3 along the line
shown in Figure \ref{W51A_rgb} (middle). The PV maps illustrate the
kinematic relationship of each radio continuum source and associated
molecular cloud in W51A. The radio continuum sources of G49.5-0.4 appear
to be associated with the molecular cloud at 58 \kms, and the G49.4-0.3
sources with a 53 \kms\ molecular component. Figure \ref{W51A_pvmap} also
shows the clumpiness of the HV stream as well as the spatial relationship
between the radio continuum sources and the molecular clouds. The HV
stream with velocity 68 \kms\ is strong in front of the G49.5-0.4 region,
while the HV stream near G49.4-0.3 is weak.

\subsubsection{W51B}

W51B consists of six radio continuum sources: G49.2-0.3, G49.17-0.21,
G49.10-0.27, G49.1-0.4, G49.0-0.3 and G48.9-0.3. In Figure \ref{W51B_rgb},
we mark the radio continuum sources listed by \cite{Koo97a} on the
three-color IRAC composite image. The molecular cloud toward the W51B
region is distributed within the velocity range from 55 to 75 \kms. Figure
\ref{W51B_rgb} (middle) shows the \co\ \jt\, intensity integrated between
55 and 75 \kms, which is coincident with the bright IRAC emission. Most
strong molecular clouds appear around the radio continuum sources. Eighty
percent of the YSO candidates (lower panel) in W51B are early stage
objects (Stage 0/I or ambiguous).

The \ion{H}{2} regions of W51B are distributed parallel to the Galactic
plane. Molecular clouds are distributed along $b \sim -0\fdg3$. To
analyze the detailed structure of each \ion{H}{2} region and associated
molecular cloud, we divide the W51B region into six groups: G49.2-0.3,
G49.17-0.21, G49.10-0.27, G49.1-0.4, G49.0-0.3 and G48.9-0.3.

Figure \ref{W51B_pv} shows the PV maps of \co\ \jt, \tco\ \jt\, and $
^{12/13}R_{2-1}$, through the three \ion{H}{2} regions in the W51B region
from B to B$^\prime$ in Figure \ref{W51B_rgb} (middle). Most CO emission
appears between 55 and 75 \kms. The molecular clouds associated with
G48.9-0.3 and G49.0-0.3 are strong near 68 \kms. The $^{12/13}R_{2-1}$
ratio around G49.2-0.3 is smaller than 1.5 (red contours) at the offset
position of $-$7\farcm5, which implies that the optical depth of the
molecular gas is large at that position. Several YSOs are identified
to the east of G49.2-0.3 ($l,b = 49\fdg26, -0\fdg34$) on Figure
\ref{W51B_rgb}. In Figure \ref{W51B_pv} (bottom), the molecular cloud
associated with these YSOs, also shows a low ratio ($ ^{12/13}R_{2-1}
< 1.5$), like the molecular cloud associated with G49.2-0.3.

\subsection{Individual Regions}
\subsubsection{G49.5-0.4} 
\label{sec:G49.5-0.4}

\begin{figure}
\epsscale{1.2}
\plotone{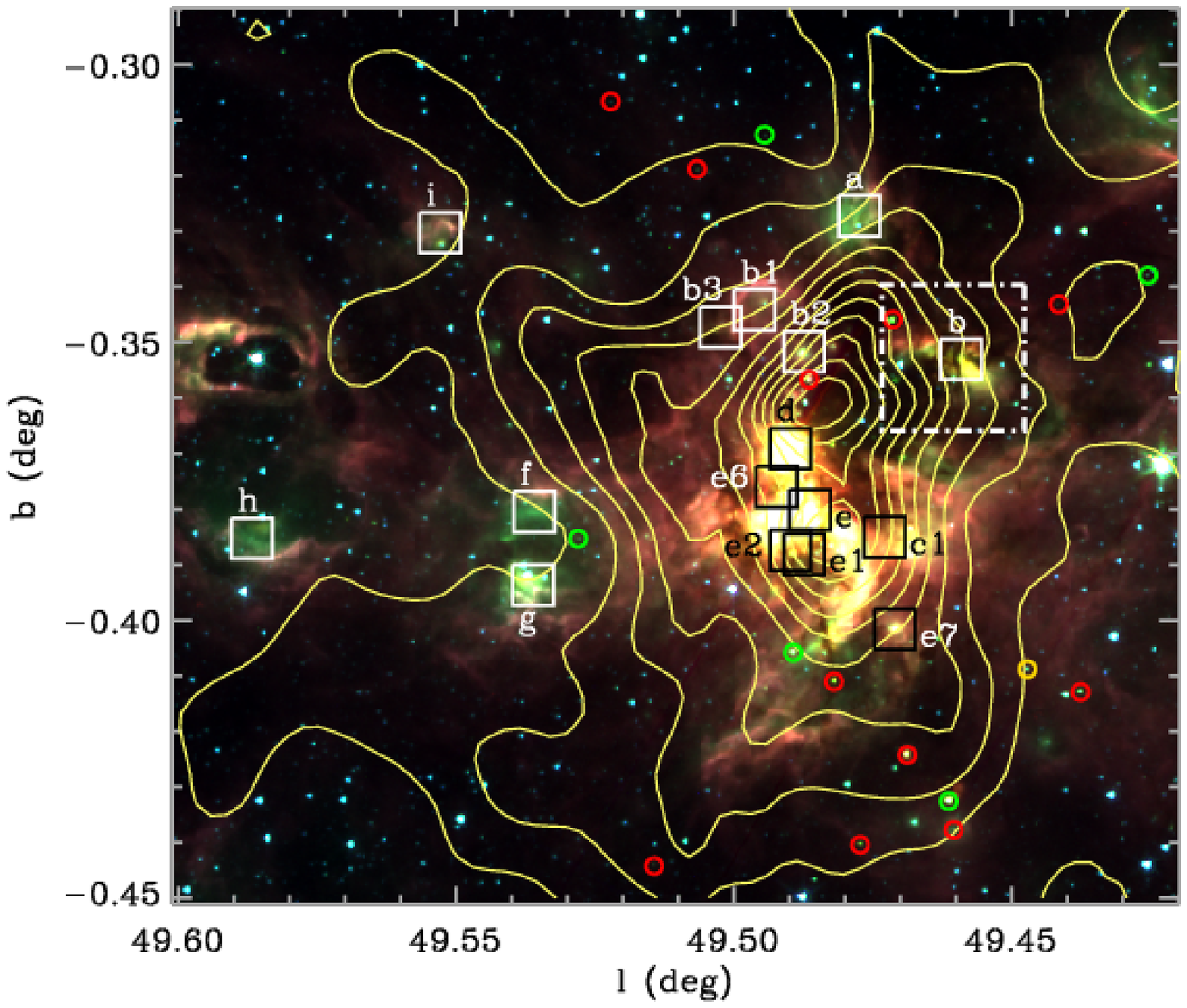}

\caption{The \tco\ \jt\ intensity integrated over the velocity range
from 45 to 75 \kms\ overlaid on the color image of the G49.5-0.4
region composed of IRAC 5.8 \micron\ (red), 4.5 \micron\ (green),
and 3.6 \micron\ (blue). Contour levels are 20, 40, 60, 80, 100,
120, 140, 160, 180, 200, 220, and 240 K \kms. Squares are the ZAMS
OB stars listed by \cite{Mehringer94}. YSO candidates (open circles)
are marked in red for Stage 0/I, yellow for Stage II, and green for
ambiguous sources. The dash-dotted rectangle shows the field in Figure
\ref{G49.5-0.4_b_spectra}.}

\label{G49.5-0.4_rgb}
\end{figure}

\begin{figure}
\epsscale{1.1}
\plotone{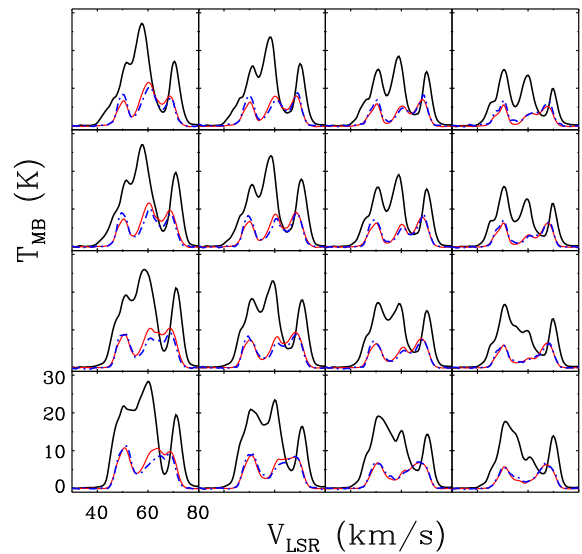}

\caption{Tile map of the \co\ \jt\ (thick solid line), \tco\
\jt\ (thin solid line), and \tco\ \jo\ (dash-dotted line) spectra
of G49.5-0.4 component b seen in the dash-dotted box in Figure
\ref{G49.5-0.4_rgb}. Spacing of the mapping grid is 22\arcsec.}

\label{G49.5-0.4_b_spectra}
\end{figure}

\begin{figure}
\epsscale{1.1}
\plotone{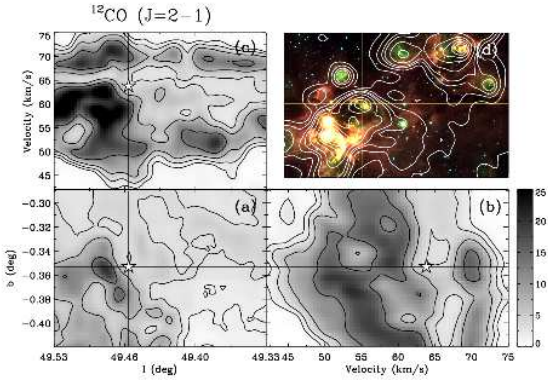}

\caption{Spatial and position-velocity diagrams of component b of
G49.5-0.4 in \co\ \jt. (a) is channel map at 64 \kms, which is RRL
velocity. The star symbol represents component b of G49.5-0.4. (b) and
(c) are position-velocity maps in latitude and longitude across the
solid lines in (a). The \co\ contour levels are 1.6, 3.2, 6.4, 12.8,
and 25.6 K in $T_{MB}$. (d) is composite image of IRAC 5.8 $\micron$
(red), 4.5 $\micron$ (green), and 3.6 $\micron$ (blue). The contours show
the 21 cm radio continuum emission \citep{Koo97}. The contour levels are
0.015, 0.05, 0.1, 0.15, 0.3, 0.6, 0.9, 1.2, 1.5, and 1.8 Jy beam$^{-1}$.}

\label{G49.5-0.4b_pv_12co}
\end{figure}

\begin{figure}
\epsscale{1.1}
\plotone{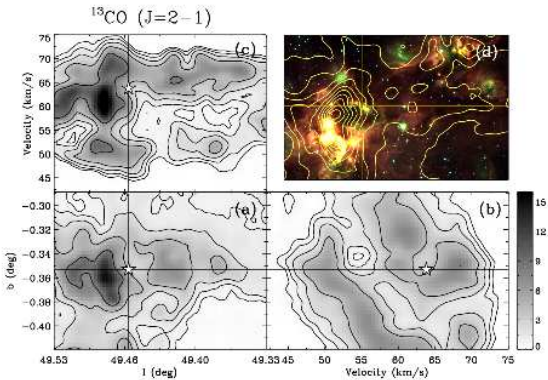}

\caption{Same as in Figure \ref{G49.5-0.4b_pv_12co}, for \tco\ \jt.
The \tco\ contour levels are 0.5, 1.0, 2.0, 4.0, 8.0, and 16.0 K
in $T_{MB}$. Contours on (d) show the \tco\ \jt\ intensity integrated
from 42 to 75 \kms.  The contour levels are 15, 45, 75, 105, 135, 165,
195, 225, 255, and 285 K \kms.}

\label{G49.5-0.4b_pv_13co}
\end{figure}

\begin{figure}
\epsscale{1.1}
\plotone{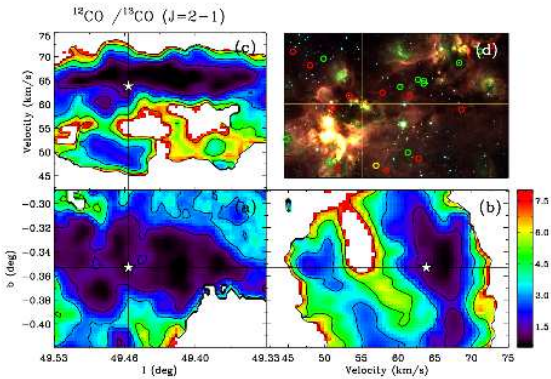}

\caption{Same as in Figure \ref{G49.5-0.4b_pv_12co}, for ratio of
$^{12/13}R_{2-1}$. Contour levels are 1.5, 3.0, 5.0, and 7.0. YSO
candidates (open circles) on (d) are marked in red for Stage 0/I, yellow
for Stage II, and green for ambiguous sources.}

\label{G49.5-0.4b_pv_r1213}
\end{figure}

Figure \ref{G49.5-0.4_rgb} shows the complicated structure of G49.5-0.4
in various wavelengths. Although the IRAC emission and the \tco\ \jt\
integrated intensity show similar structure, the peak position of the
\tco\ \jt\ intensity coincides with a dark part of the IRAC composite
image near the southeast region of component b. Some radio sources are
located in the brightest part of the molecular cloud. Two bright radio
continuum sources, W51d and W51e, coincide with the infrared sources IRS
2 and IRS 1 and are adjacent to two peaks of the integrated intensity
map of the \tco\ \jt\ emission. Other sources (a, b, f, g, h, and i) are
located near the edge of the molecular cloud. These radio sources are
coincident with the bright region in the IRAC 4.5 \micron\ band
(green) probably indicative of shocked H$_2$.

The spatial distribution of the radio continuum sources and associated
molecular clouds can be constructed by comparing the velocity of ionized
gas, CO emission, and absorption lines. The average velocity of the
ionized gas is 59 \kms, which was measured using the H109$\alpha$
recombination line by \cite{Wilson70}. CO emission is brightest
at $V_{LSR}$ = 61.0 \kms\ (Figure \ref{ave_spectra}). The ionized
and molecular gas have similar velocities, which implies that the
molecular cloud at this velocity is associated with the radio continuum
source. Toward G49.5-0.4, the central velocities of the ionized gas
vary from 48 to 77 \kms\ \citep{Mehringer94}. H$_{2}$CO absorption has
been detected between 52 and 73 \kms\ \citep{Arnal85}. \ion{H}{1}
21 cm absorption has been observed at 50, 62, and 69 \kms\
\citep{Koo97a}. H$_{2}$CO absorption between 66 and 70 \kms\ and
\ion{H}{1} absorption at 69 \kms\ implies that the HV stream is in
front of G49.5-0.4. Regarding the spatial location of each radio
continuum source in the G49.5-0.4 region, \cite{Arnal85} suggested
that components a, b, and c are closer to the observer than components
d and e by comparing the H$\alpha$109 RRL and H$_{2}$CO absorption
spectra. \cite{Koo97a} reported that the \ion{H}{1} cloud with 62 \kms\
velocity associated with G49.5-0.4, and components a, b, and e are
either embedded in or behind this cloud while components f and g are
located either in front of or the outside the cloud.  Also, all radio
continuum sources in the G49.5-0.4 region are spatially close because all
\ion{H}{1} spectra detected toward these sources have similar absorption
profiles. Most of the bright radio components of G49.5-0.4 in the W51
\ion{H}{2} region complex must be directly associated with the molecular
clouds at 60 \kms.

CO spectra in the G49.5-0.4 region show complex structure. The
average \co\ \jt\, spectrum of G49.5-0.4 in Figure \ref{ave_spectra}
is divided into two components with velocities of 59 and 68 \kms\
at the emission peaks. The brighter component between 45 to 65 \kms\
is associated with the radio continuum sources in G49.5-0.4. Figure
\ref{G49.5-0.4_b_spectra} shows the CO spectra around component b of
G49.5-0.4. There are three peaks in the \co\ \jt\ emission at 52, 58, and
70 \kms. The intensities of \tco\ \jt\ and \jo\ at 66 \kms\ are greater
than \co\ \jt\ intensities, which implies self-absorption in the \co\
\jt\ emission. These CO self-absorption features clearly appear as a
low ratio of $^{12/13}R_{2-1}$ on PV maps inside the first contour ($<
1.5$) on Figure \ref{W51A_pvmap} (bottom).

\begin{figure}
\epsscale{1.2}
\plotone{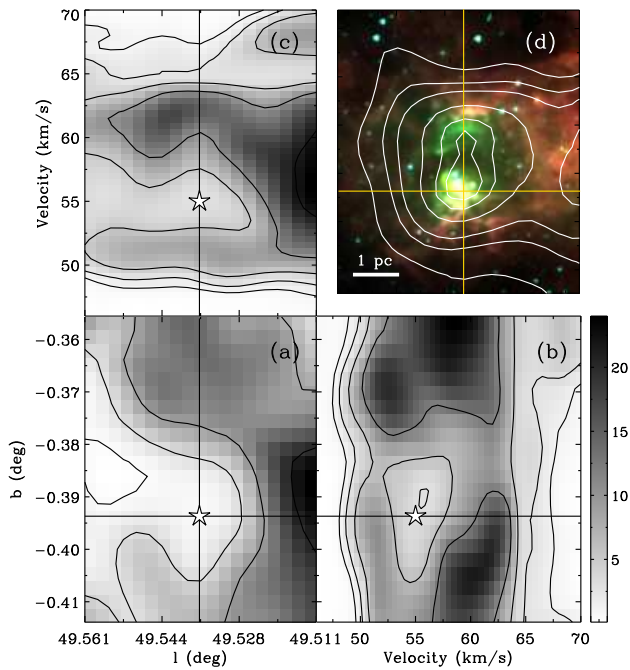}

\caption{Spatial and position-velocity diagrams of component g of
G49.5-0.4 in \co\ \jt. (a) is the channel map at 55 \kms. The star
symbol represents the component g. Contour levels are 1.5, 3, 6, and 12
K in $T_{MB}$. (b) and (c) are position-velocity maps in latitude and
longitude across the solid lines in (a). (d) is the composite image
of IRAC 5.8 \micron\ (red), 4.5 \micron\ (green), and 3.6 \micron\
(blue). The contours show the 21 cm radio continuum emission \citep
{Koo97}. The contour levels are 0.01, 0.03, 0.05, 0.10, 0.15, and 0.18
Jy beam$^{-1}$. Scale bar in the bottom left corner represents 1 pc at
the distance of 6 kpc.\\}

\label{G49.5-0.4g_pvmap}
\end{figure}

\begin{figure}
\epsscale{1.2}
\plotone{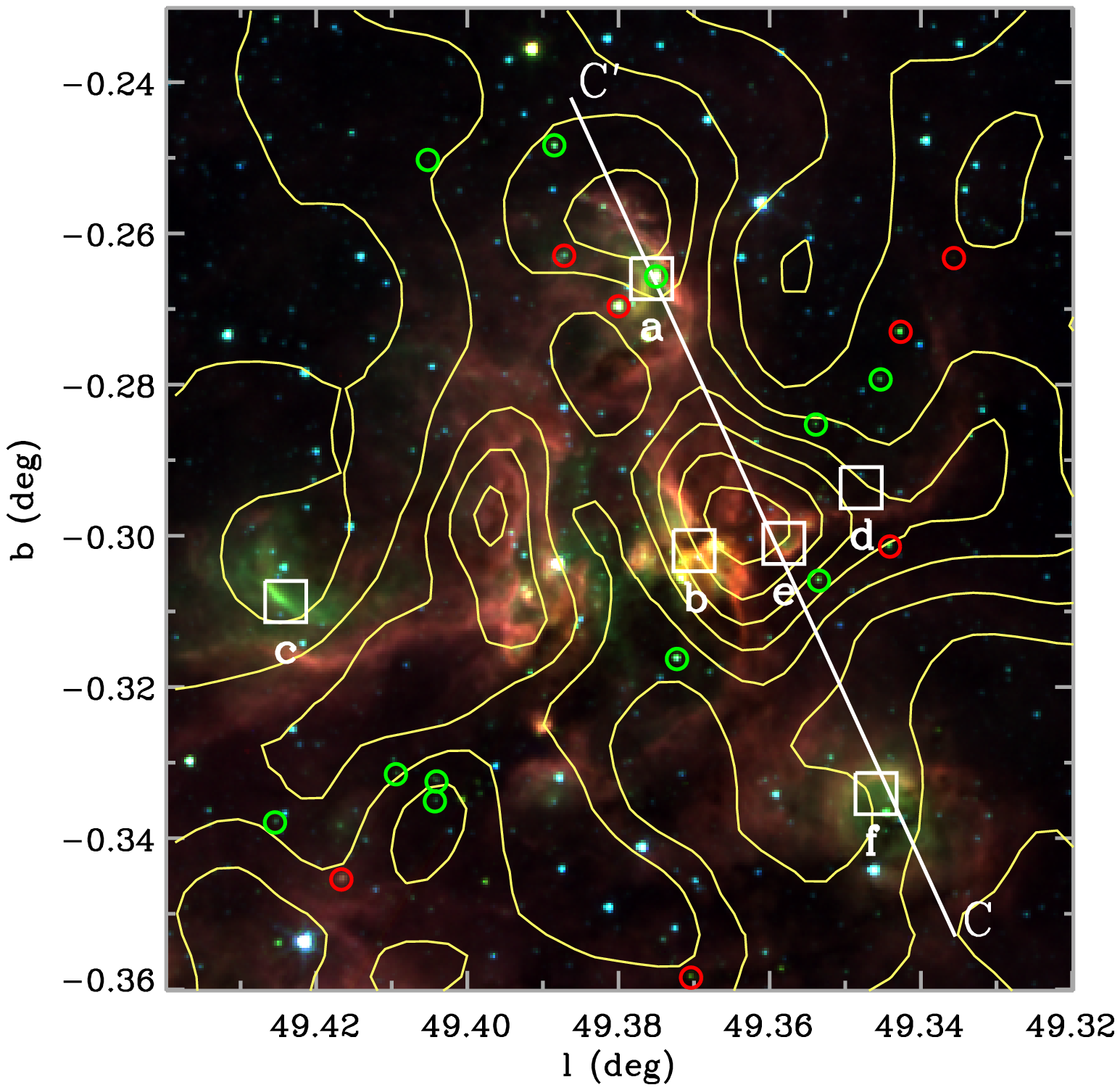}

\caption{The \co\ \jt\ intensity integrated over the velocity range
from 45 to 55 \kms\ overlaid on the color image of the G49.4-0.3 region
composed of IRAC 5.8 \micron\ (red), 4.5 \micron\ (green), and 3.6
\micron\ (blue). Contour levels are 25, 50, 75, 100, 125, 150, and 175
K \kms. Squares are the ZAMS OB stars listed by \cite{Mehringer94}. YSO
candidates (open circles) are marked in red for Stage 0/I, yellow for
Stage II, and green for ambiguous sources. The oblique line shows the cut
for the position-velocity map presented in Figure \ref{G49.4-0.3_sn_pv}.}

\label{G49.4-0.3_rgb}
\end{figure}

\begin{figure}
\epsscale{1.2}
\plotone{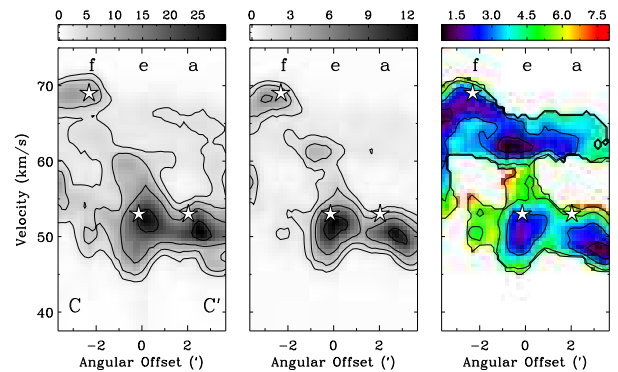}

\caption{\co\ \jt\ ({\it left}), \tco\ \jt\ ({\it middle}) and
$^{12/13}R_{2-1}$ ({\it right}) position-velocity map through the radio
continuum sources a, e, and f of G49.4-0.3 from southeast (C) to northwest
(C$^\prime$) of the map in Figure \ref{G49.4-0.3_rgb}. The temperature
scale is indicated on the top (K in $T_{MB}$). Contour levels of the
\co\ \jt\ and \tco\ \jt\ maps are 10, 20, 40, and 80\% of peak value
in the map. Contour levels of the $^{12/13}R_{2-1}$ map are 1.5, 3.0,
5.0, and 7.0. Star symbols represent the position of the radio continuum
sources. The velocity of a and e is the mean velocity of the ionized gas
\citep{Pankonin79}. The velocity of component f comes from the velocity
of the molecular cloud around the radio continuum source.}

\label{G49.4-0.3_sn_pv}
\end{figure}

\cite{Arnal85} proposed that component b of G49.5-0.4 is an \ion{H}{2}
regions created by O stars formed in the collision between the HV stream
and the molecular cloud at 61-63 \kms. The RRL velocity of component b has
been detected at 63.8 \kms\ in H109$\alpha$ \citep{vanGorkom80}. Figures
\ref{G49.5-0.4b_pv_12co}--\ref{G49.5-0.4b_pv_r1213} show the
spatial and kinematic structure around component b of G49.5-0.4
in \co, \tco\ \jt, and $^{12/13}R_{2-1}$. As seen in Figures
\ref{G49.5-0.4b_pv_12co}--\ref{G49.5-0.4b_pv_r1213}, component b lies
between the main molecular cloud and the HV stream on the PV map. A
clear difference between the PV maps of each CO isotopologue appears at
66 \kms. The \co\ \jt\, intensity is very weak at the velocity between
the main molecular cloud and the HV stream. However, the PV map of
\tco\ \jt\, does not show an apparent boundary at this velocity so that
$^{12/13}R_{2-1}$ becomes small near 66 \kms. A deep absorption feature
of H$_{2}$CO spectra \citep{Fomalont73} and lower opacity molecular
species such as CS $J=3-2$ \citep{Penzias71} and HCN \citep{Snyder71}
have been detected at 66 \kms, which imply the presence of cold dense
material in this region. In Figure \ref{G49.5-0.4b_pv_r1213}(d) there
are dark filamentary structures on the IRAC composite image. Several YSOs
in very early stages are detected in the dark filaments which coincide
with the inner area with $^{12/13}R_{2-1}$ of 1.5 (the first contour in
Figure \ref{G49.5-0.4b_pv_r1213}(a)).

Component g of G49.5-0.4 was observed at 54 \kms\ in [\ion{S}{3}]
emission line \citep{Goudis82}. The average H$\alpha$ velocity is
54.6 $\pm$ 13.7 \kms\ \citep{Crampton78}. To investigate the velocity
structure around component g, we present in Figure \ref{G49.5-0.4g_pvmap}
the channel map at 55 \kms, which is the ionized gas velocity, and the
position velocity map across component g. A prominent molecular gas cavity
exists around component g in Figure \ref{G49.5-0.4g_pvmap}(a). This cavity
appears at a velocity of 55 \kms\ of both ($b,v$) (b) and ($l,v$) (c)
maps. Figure \ref{G49.5-0.4g_pvmap}(c) shows the velocity gradient from
the expansion of molecular clouds around component g. In addition to the
velocity structure, the green color in Figure \ref{G49.5-0.4g_pvmap}(d)
caused by the shocked H$_{2}$ emission is a strong evidence for the
interaction of the ionizing star(s) and surrounding molecular cloud.

\subsubsection{G49.4-0.3} 
\label{sec:G49.4-0.3}

G49.4-0.3 is located to the northwest of G49.5-0.4 and is composed of six
subcomponents \citep{Mehringer94}. Figure \ref{G49.4-0.3_rgb} shows the
detailed structure of G49.4-0.3 in \co\ \jt\ and IRAC bands. The bright
emission in IRAC bands coincides with the \co\ \jt\ intensity. Shocked
bright 4.5 \micron\ emission (green) appears around most radio continuum
sources, except component d.

At least three different velocity components are shown in the average
spectra of G49.4-0.3 (Figure \ref{ave_spectra}). The brightest component
at 52 \kms\ is associated with most of the radio continuum sources in
G49.4-0.3 because the mean velocity of the ionized gas is at 53 \kms\
\citep{Pankonin79}. However, the velocities of the individual components
have not been measured. The H$_{2}$CO absorption for components b and
c is observed at 51 and 63 \kms\ \citep{Arnal85}, and the \ion{H}{1}
absorption for components a, b, c, d, and f has velocities between 42
and 66 \kms\ \citep{Koo97a}.

Figure \ref{G49.4-0.3_sn_pv} shows the PV map through components a,
e, and f. In the \tco\ \jt\ PV map, there are three different velocity
components. The brightest molecular cloud lies at 52 \kms, another at
62 \kms\ and the other at 69 \kms. All radio sources, except component
f, are coincident with the molecular clouds at 52 \kms. The \ion{H}{1}
absorption at 53 \kms\ is dominant toward components a, b, c, and d, but
for component f it is negligible \citep{Koo97a}. Taking into account the
dominant \ion{H}{1} absorption and CO molecular cloud at the velocity
of 69 \kms, component f is likely to be associated with the molecular
cloud at 69 \kms.

\begin{figure}
\epsscale{1.2}
\plotone{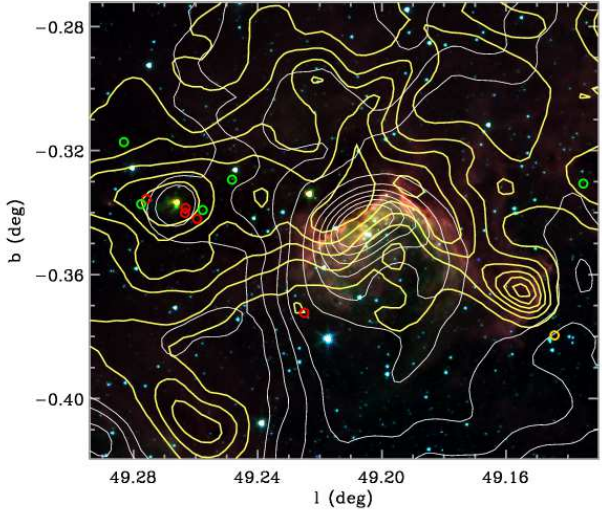}

\caption{The \tco\ \jt\ intensity integrated over the velocity range
from 57 to 76 \kms\ overlaid on the color image of the G49.2-0.3 region
composed of IRAC 5.8 \micron\ (red), 4.5 \micron\ (green), and 3.6
\micron\ (blue). Thick contour levels are 7, 14, 21, 28, 35, and 42
K \kms. Thin white contours show 21 cm radio continuum map. Contour
levels are 0.015, 0.03, 0.06, 0.1, 0.3, 0.6, 0.9, 1.2, 1.5, and 1.8 Jy
beam$^{-1}$ \citep{Koo97}. YSO candidates (open circles) are marked in
red for Stage 0/I, yellow for Stage II, and green for ambiguous sources.\\}

\label{G49.2-0.3_rgb}
\end{figure}

\subsubsection{G49.2-0.3} 
\label{sec:G49.2-0.3}

Figure \ref{G49.2-0.3_rgb} shows G49.2-0.3, which is the largest and
brightest \ion{H}{2} region in W51B, in IRAC bands as a three-color
composite image, \co\ \jt\ by thick yellow contours, and 21 cm radio
continuum by thin white contours. The radio continuum emission of
G49.2-0.3 is apparently coincident with the bright region in IRAC
bands. The 21 cm continuum brightness decreases rapidly toward the
molecular cloud. The \co\ \jt\ molecular cloud emission integrated between
57 and 76 \kms\ encompasses the half of the 21 cm radio continuum source
toward the north. The morphology of the radio continuum source and
the distribution of the CO emission indicate that G49.2-0.3 is a good
example of blister type \ion{H}{2} region \citep{Israel78}. The RRL
velocity of the \ion{H}{2} region is 67.6 \kms\ observed in H109$\alpha$
by \cite{Pankonin79} and the peak velocity of the \tco\ \jt\ emission
averaged over all pixels shown in Figure \ref{G49.2-0.3_rgb} is 67.5 \kms\
(Figure \ref{ave_spectra}). The \ion{H}{2} region G49.2-0.3 forms on the
boundary of the molecular cloud at 67.5 \kms. The \tco\ \jt\ emission is
strong in the east side of the \ion{H}{2} region where many YSOs appear
around ($l,b$ = 49\fdg26, $-$0\fdg34). The peak velocity of \tco\ \jt\
emission is 68 \kms, which is close to the RRL velocity of G49.2-0.3. As
discussed in the large scale structure of W51B, detailed velocity
structure of CO lines through G49.2-0.3 and YSO clustering region is
shown in Figure \ref{W51B_pv}. These two regions divide into two different
clumps on the \tco\ \jt\ PV map of Figure \ref{W51B_pv} ({\it middle})
and the integrated intensity map of Figure \ref{G49.2-0.3_rgb}. The ratios
of $^{12/13}R_{2-1}$ are smaller than 1.5 at the position of \ion{H}{2}
region and the YSO clustering region around 66 \kms\ (Figure \ref{W51B_pv}
{\it bottom}). The regions with the small ratios are associated with
the star-forming activity. 

\begin{figure}
\epsscale{1.2}
\plotone{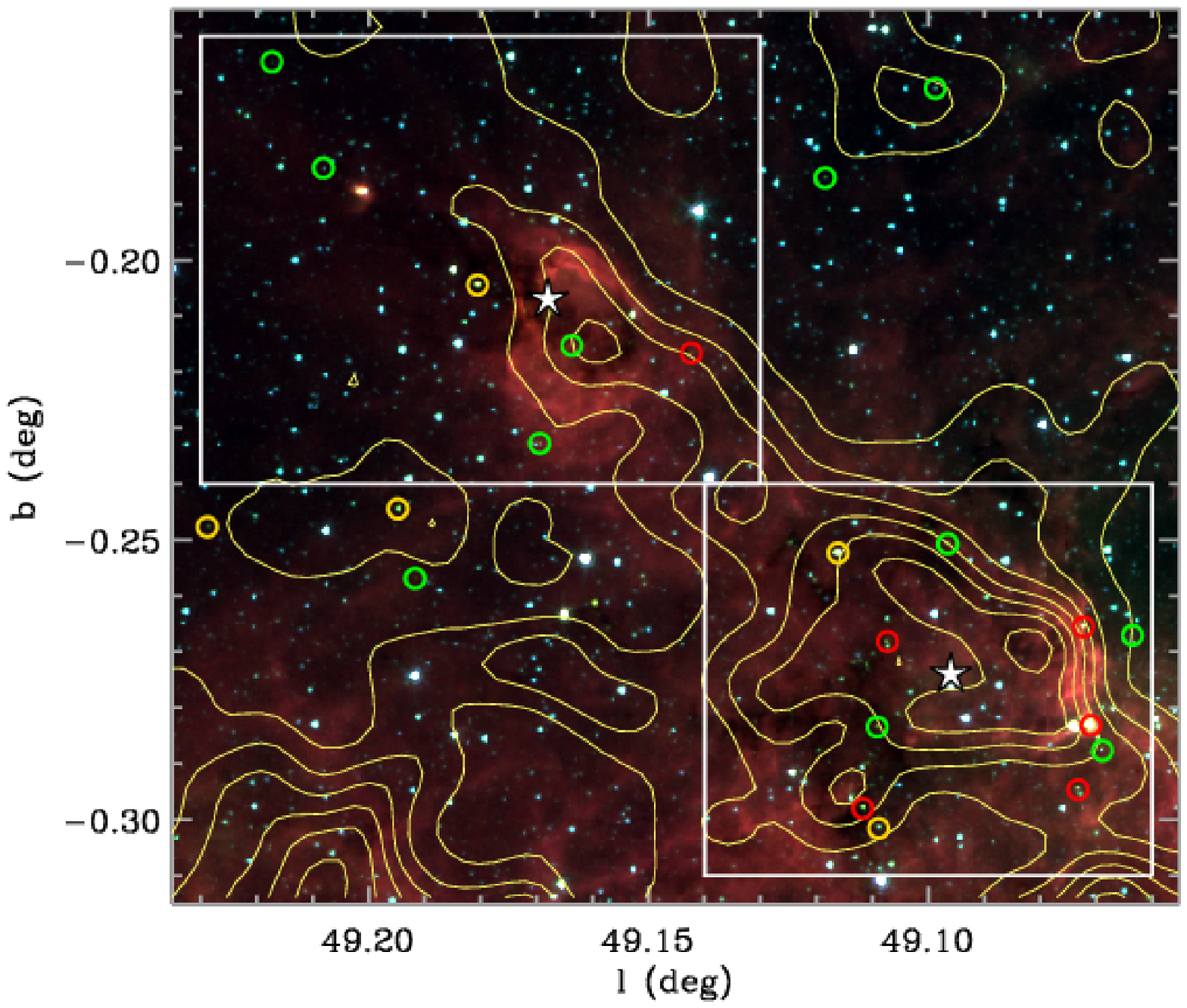}

\caption{The \co\ \jt\ intensity integrated over the velocity range
from 55 to 75 \kms\ overlaid on the color image of the G49.17-0.21 and
G49.10-0.27 region composed of IRAC 5.8 \micron\ (red), 4.5 \micron\
(green), and 3.6 \micron\ (blue). Contour levels are  20, 40, 60, 80,
100, 120, and 140 K \kms. Star symbols are the radio continuum sources
\citep{Koo97a}. YSO candidates (open circles) are marked in red for
Stage 0/I, yellow for Stage II, and green for ambiguous sources.\\}

\label{G49.1-0.2_rgb}
\end{figure}

\begin{figure}
\epsscale{1.2}
\plotone{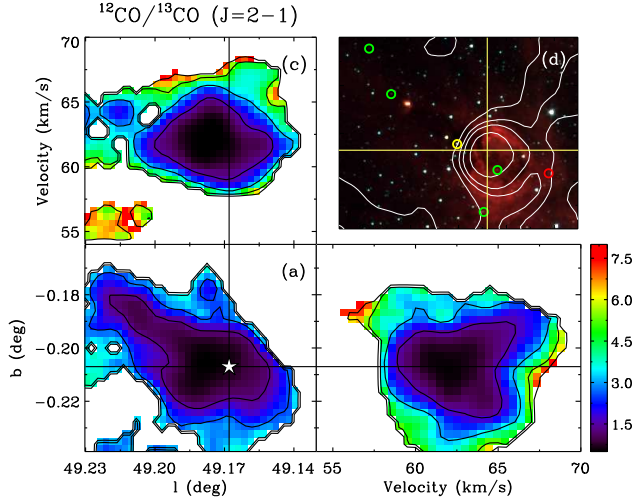}

\caption{Spatial and position-velocity diagrams of G49.17-0.21 in
$^{12/13}R_{2-1}$. (a) is channel map at 61 \kms. (b) and (c) are
position-velocity maps in latitude and longitude across the solid lines in
(a). Contour levels are 1.5, 3.0, 5.0, and 7.0. (d) is composite image of
IRAC 5.8 $\micron$ (red), 4.5 $\micron$ (green), and 3.6 $\micron$ (blue).
The contours show the 21 cm radio continuum emission \citep{Koo97}. The
contour levels are 0.01, 0.03, 0.05, and 0.10 Jy beam$^{-1}$. The star
symbol is the radio continuum source \citep{Koo97a}. YSO candidates
(open circles) on (d) are marked in red for Stage 0/I, yellow for Stage
II, and green for ambiguous sources.\\}

\label{G49.17-0.21_pv}
\end{figure}

\subsubsection{G49.17-0.21 and G49.10-0.27} 
\label{sec:G49.17-0.21 and G49.10-0.27}

G49.17-0.21 and G49.10-0.27 are faint in IRAC bands, compared with other
radio continuum sources in the W51B region (Figure \ref{W51B_rgb}).
Figure \ref{G49.1-0.2_rgb} shows the \co\ \jt\ intensity superposed
on the IRAC three-color composite image. G49.17-0.21 is located in the
northeast of the map. A cometary shape that consists of a head around
the \ion{H}{2} region and a tail toward northeast appears in all four
IRAC bands as a dark feature. The \co\ \jt\ integrated intensity map
covers the bright IRAC emission around the \ion{H}{2} regions over
a large scale.  The brightest CO emission coincides with the darker
part of the IRAC image. In Figure \ref{G49.17-0.21_pv}, we present the
ratio of $^{12/13}R_{2-1}$ on the channel map of 61 \kms\ and PV maps
of G49.17-0.21. The region inside the first contour ($^{12/13}R_{2-1}
< 1.5$) coincides with the dark part of the IRAC image, where YSOs
lie. The small ratio of $^{12/13}R_{2-1}$ implies that the molecular
cloud around G49.17-0.21 is exhibiting self-absorption in \co\ \jt. In
the case of G49.10-0.27, a dark feature on the IRAC image appears around
the \ion{H}{2} region, where many YSOs are identified. The RRL velocity
of 61 \kms\ is measured only for G49.10-0.27 by \cite{Pankonin79}.

\begin{figure}
\plotone{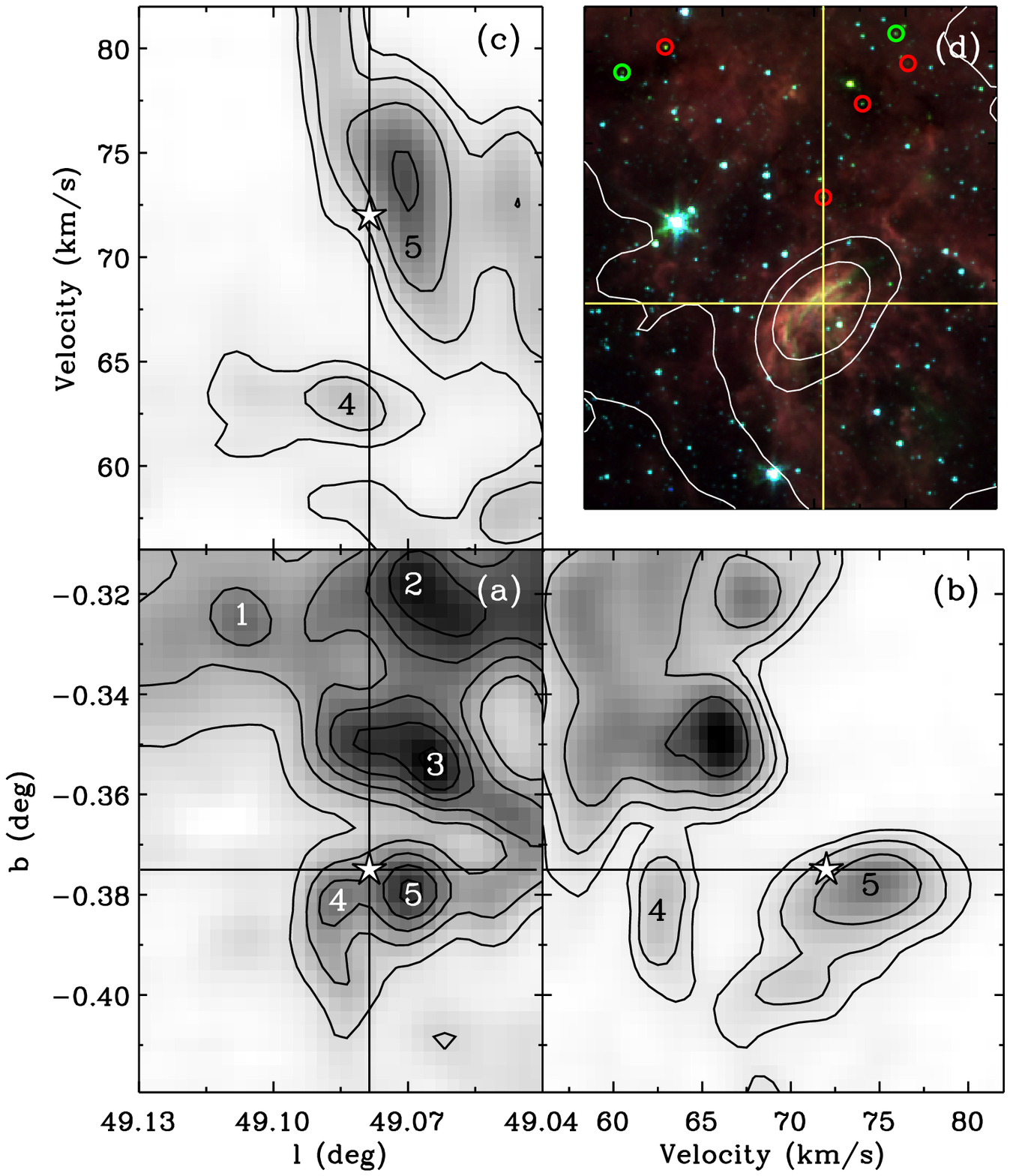}

\caption{Spatial and position-velocity diagrams of G49.1-0.4 in \co\
\jt. (a) is the \co\ \jt\ intensity integrated over the velocity
range between 56 and 82 \kms. Contour levels are 30, 60, 90, 120, 150,
and 180 K \kms. The grayscale is a stretch from 0 to 190 K \kms. Star
symbol is the radio continuum source \citep{Koo97a}. (b) and (c) are
position-velocity maps in latitude and longitude across the solid lines
in (a). Contour levels are 1.6, 3.2, 6.4, 13.0 K.  The grayscale is a
stretch from 0 to 20 K. (d) is the composite image of IRAC 5.8 \micron\
(red), 4.5 \micron\ (green), and 3.6 \micron\ (blue). The contours show
the 21 cm radio continuum emission \citep{Koo97}. The contour levels are
0.05, 0.10, 0.30, and 0.50 Jy beam$^{-1}$. YSO candidates (open circles)
are marked in red for Stage 0/I, and green for ambiguous sources.\\}

\label{G49.1-0.4_coint_bv_lv_rgb}
\end{figure}
\begin{figure}
\plotone{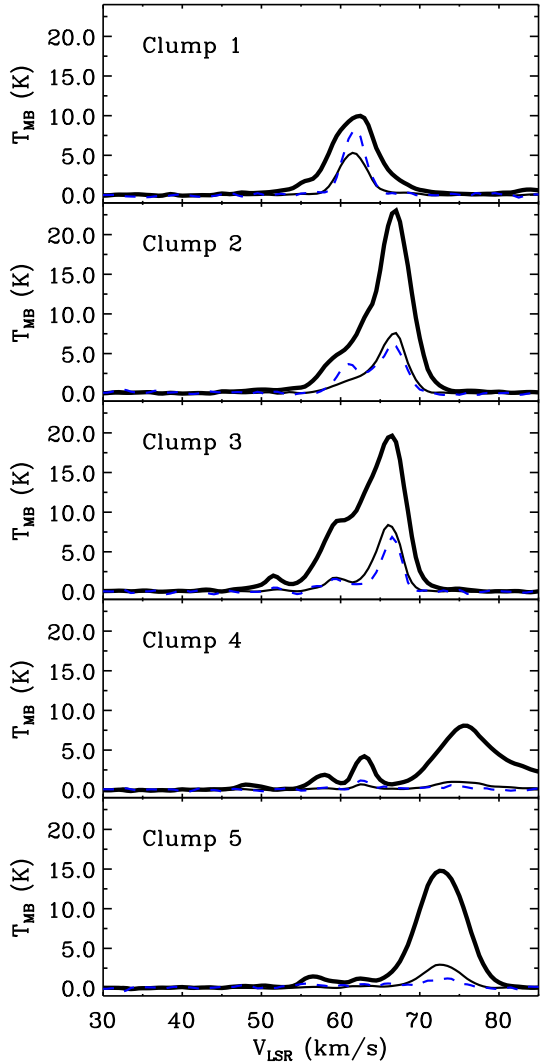}

\caption{The \co\ \jt\ (thick solid line), \tco\ \jt\ (thin solid
line), and \tco\ \jo\ (dash-dotted line) spectra at the peak position
of each clump in Figure \ref{G49.1-0.4_coint_bv_lv_rgb}(a). The 75 \kms\
component in the fourth panel actually belongs to clump 5.}

\label{G49.1-0.4_ave_spectra}
\end{figure}

\subsubsection{G49.1-0.4} 
\label{sec:G49.1-0.4}

Figure \ref{G49.1-0.4_coint_bv_lv_rgb} shows the spatial and kinematic
structure of the molecular cloud around G49.1-0.4. The molecular cloud
consists of five clumps on the \co\ \jt\ integrated intensity map (Figure
\ref{G49.1-0.4_coint_bv_lv_rgb}(a)). Figure \ref{G49.1-0.4_ave_spectra}
shows the CO spectra at the peak positions of each clump in \tco\ \jt.
Clump 1 has a \tco\ \jt\ peak velocity of $\sim$ 62 \kms. Clumps 2
and 3 have a \tco\ \jt\ peak velocity of $\sim$ 66 \kms, which is
similar to that of the H166$\alpha$ RRL measured from G49.1-0.4 by
\cite{Pankonin79}. CO emission of clumps 1, 2, and 3 is associated with
the dark region on the IRAC composite image and there are several YSOs
in the early stage of evolution. Near the radio continuum source, there
are two clumps: clump 4 refers to the 62 \kms\ component and clump 5
refers to the 75 \kms\ component.

It is significant that the \tco\ \jt\ intensity is stronger than
the \tco\ \jo\ in clump 5 ($^{13}R_{2-1/1-0} > 2$). The high ratio of
$^{13}R_{2-1/1-0}$ means that clump 5 is significantly different from the
other clumps. Clump 5 seen in Figure \ref{G49.1-0.4_coint_bv_lv_rgb}(a)
coincides with the bright IRAC composite emission and the 21 cm radio
continuum emission seen in Figure \ref{G49.1-0.4_coint_bv_lv_rgb}(d).
The RRL velocities in G49.1-0.4 are measured at 72.4 (H109${\alpha}$),
70.9 (H109${\alpha}$), 71.4 (H137${\beta}$) and 68 (H109${\alpha}$)
\kms\ \citep{Wilson70, Pankonin79, Downes80}. Assuming the velocity of
ionized gas is 72 \kms, clump 5 is associated with G49.1-0.4 (Figure
\ref{G49.1-0.4_coint_bv_lv_rgb}(b) and (c)). Comparing the ratios
($^{12/13}R_{2-1}$ and $^{13}R_{2-1/1-0}$) at the \tco\ \jt\ peak of
clumps 3 and 5, the excitation temperature of clump 5 is about three times
higher than that of clump 3. The total mass of clump 5 is estimated to
be $2.3\times10^3$ \msun\ by a large velocity gradient (LVG) analysis
using \co\ and \tco\ \jt\ lines.

\begin{figure}
\epsscale{1.2}
\plotone{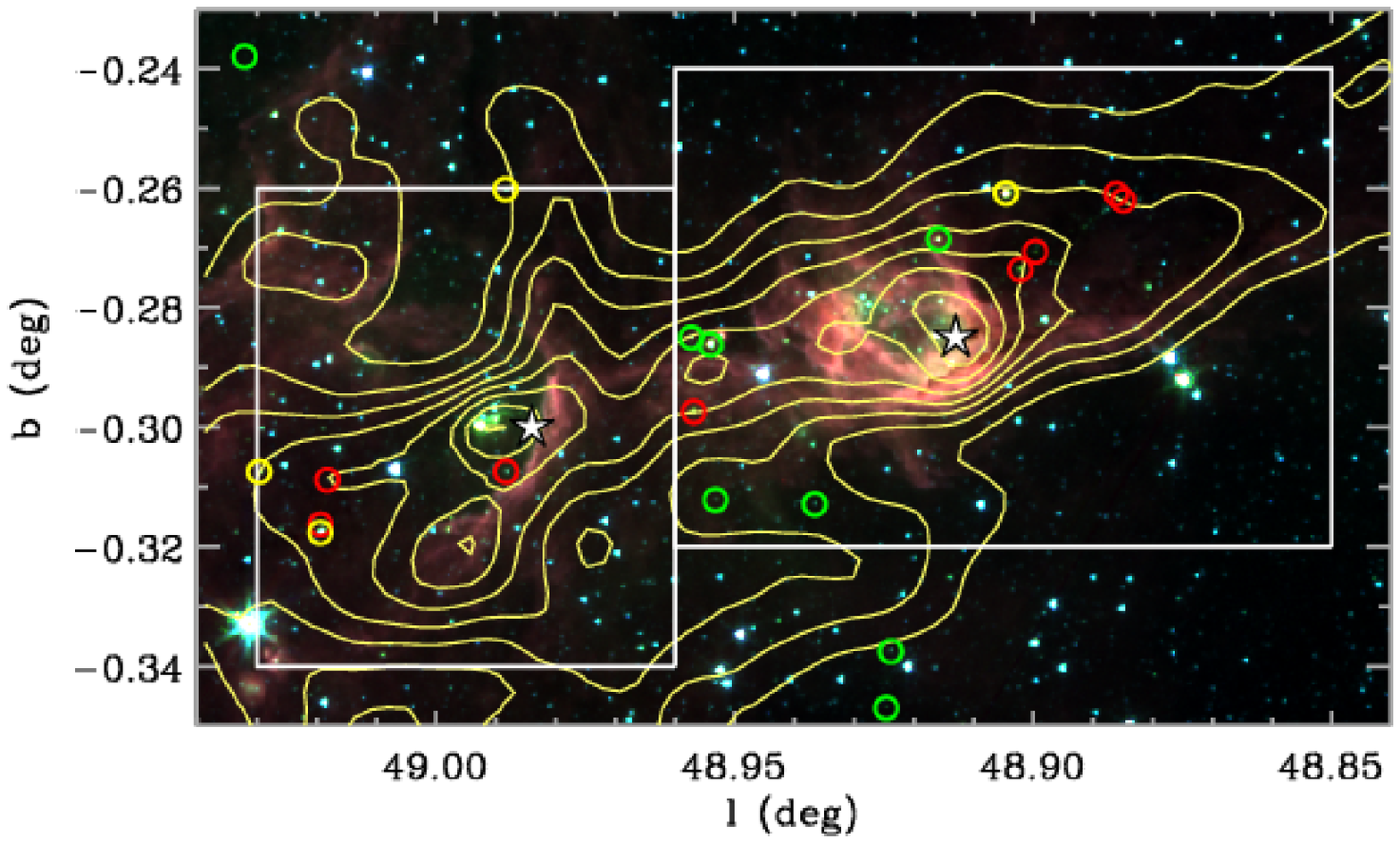}

\caption{The \co\ \jt\ intensity integrated over the velocity range
between 57 and 75 \kms\ overlaid on the color image composed of IRAC 5.8
$\micron$ (red), 4.5 \micron\ (green), and 3.6 \micron\ (blue) around
the G49.0-0.3 and 48.9-0.3 region. Contour levels are 36, 72, 108, 144,
180, 216, and 252 K \kms. Star symbols mark the radio continuum sources
\citep{Koo97a}. YSO candidates (open circles) are marked in red for
Stage 0/I, yellow for Stage II, and green for ambiguous sources.\\}

\label{G49.0_48.9-0.3_rgb_co}
\end{figure}
\subsubsection{G49.0-0.3 and G48.9-0.3} 
\label{sec:G49.0-0.3 and G48.9-0.3}

G49.0-0.3 and G48.9-0.3 are located at the west edge of the W51B region
(Figure \ref{W51B_rgb}). Figure \ref{G49.0_48.9-0.3_rgb_co} shows the \co\
\jt\ emission superposed on the IRAC three color composite image. IRAC
emission is very bright around the two radio continuum sources. The bright
IRAC emission around G49.0-0.3 makes a bow shape, which has been observed
as a faint K-band nebulosity \citep{Kumar04}. Many stellar components
seen as white spots are to the east of the bright bow. \cite{Kumar04}
found that G48.9-0.3 is composed of two sub-clusters in near-infrared
J, H, K observations. These two sub-clusters also appear in the IRAC
three-color composite image.

The average CO spectra of G49.0-0.3 and G48.9-0.3 in Figure
\ref{ave_spectra} have a similar shape. Velocities of G49.0-0.3 and
G48.9-0.3 at the \tco\ \jt\ peak intensities are 67.5 and 68.5 \kms,
respectively. The velocities of the ionized gas are measured at 64.9
and 64.4 \kms\ for G49.0-0.3 and G48.9-0.3 in the H166$\alpha$ line
\citep{Pankonin79}. The ratios of $^{13}R_{2-1/1-0}$ of G49.0-0.3 and
G48.9-0.3 are large ($>$ 0.7) around the RRL velocity, which indicates
that the molecular clouds associated with the radio continuum sources
are highly excited.

\begin{figure}
\epsscale{1.2}
\plotone{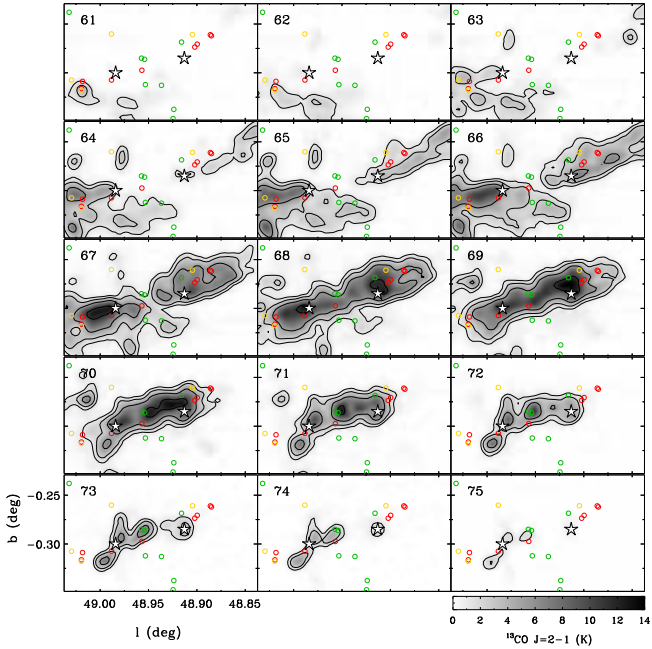}

\caption{Channel maps of G48.9-0.3 and G49.0-0.3 in the \tco\ \jt\
emission over the velocity range from 61 to 75 \kms. Contour levels
are 1.3, 2.6, 5.2, and 10.4 K in $T_{MB}$. Star symbols are the
radio continuum sources \citep{Koo97a}. YSO candidates (open circles)
are marked in red for Stage 0/I, yellow for Stage II, and green for
ambiguous sources.\\}

\label{G49.0_48.9-0.3_chanmap}
\end{figure}

Many YSOs are found in the relatively dark region of the IRAC image
along the filamentary structure of the molecular cloud. Two bright CO
components are coincident with two radio continuum sources. We confirm
that the dense regions are associated with the newly forming YSOs by
comparing the distribution of YSOs and the \tco\ \jt\ intensity on the
channel maps. Figure \ref{G49.0_48.9-0.3_chanmap} shows the \tco\ \jt\
intensity in the velocity range from 61 to 75 \kms\ per 1 \kms . There are
three groups of YSOs: one to the east of G49.0-0.3, another to the west of
G48.9-0.3, and the other between G49.0-0.3 and G48.9-0.3. Dense molecular
clouds associated with each group lie in different velocity ranges. YSOs
 between G49.0-0.3 and G48.9-0.3 are coincident with the high column
density region from the 68 \kms\ channel. As seen in the PV diagram
of \tco\ \jt\ in Figure \ref{W51B_pv} (middle), there are three dense
molecular clumps around G49.0-0.3 and G48.9-0.3. While the two clumps
located on either side of the whole cloud are associated with radio
continuum G49.0-0.3 and G48.9-0.3, the molecular clump between the two
\ion{H}{2} regions lies toward higher velocity. These two \ion{H}{2}
regions form an expanding arc of gas in PV diagram \citep{Moon98}
(Figure \ref{W51B_pv}).

\section{Discussion}
\label{co:discussion}

\subsection{Spatial Distribution of the Gas, YSOs, and Radio Continuum
Sources}

\begin{figure}
\epsscale{1.2}
\plotone{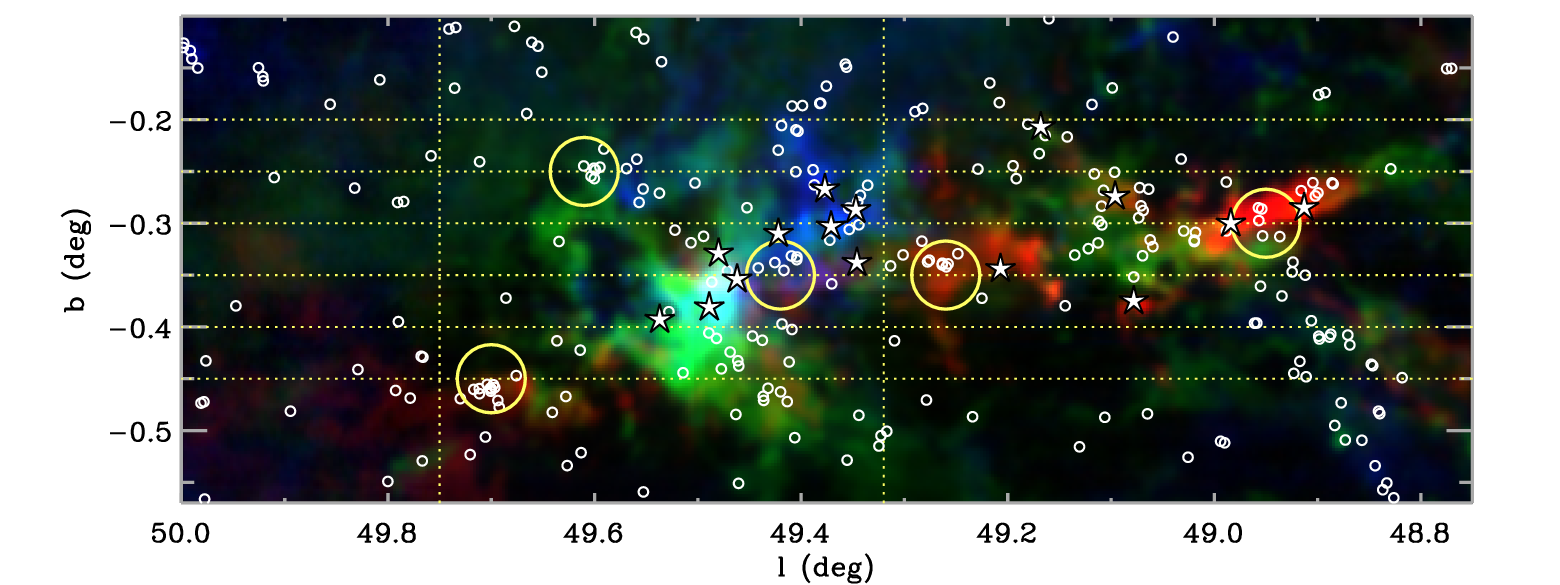}

\caption{Spatial distribution of the \ion{H}{2} regions (stars) and YSOs
(small open circles) in the W51 complex, overplotted on a color image
composed of \co\ \jt\ intensity integrated over velocity range 30--55
\kms\ (blue), 56--65 \kms\ (green), and 66--85 \kms\ (red). The dotted
horizontal lines are guide lines by 0\fdg05 step. Two vertical dotted
lines are guide lines for dividing W51A from W51B. Large thick open
circles represent the YSO clustering regions.\\}

\label{w51_small}
\end{figure}

\begin{figure}
\epsscale{1.14}
\plotone{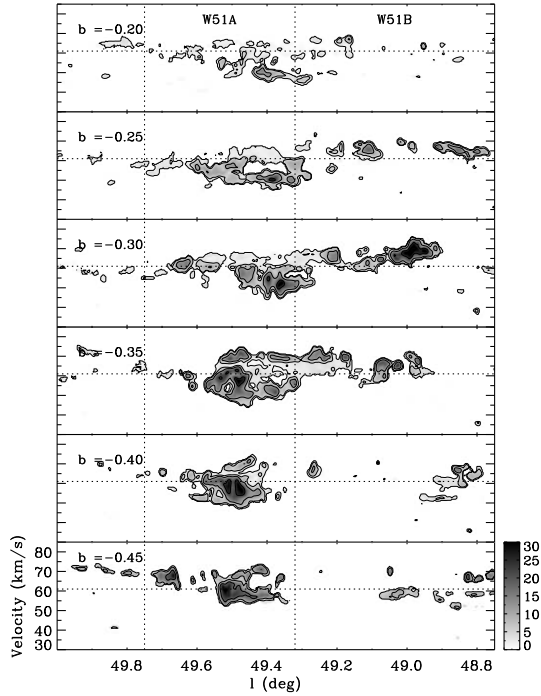}

\caption{The longitude-velocity map of \co\ \jt\ at fixed latitudes
from $b=-0\fdg20$ to $b=-0\fdg45$ by 0\fdg05 step. Contour levels are 3,
6, 12, and 24 K in $T_{MB}$. Two dotted vertical lines are guide lines for
dividing the W51A and W51B regions. A dotted horizontal line shows the
boundary of the HV streaming gas.}

\label{lvmap_12co}
\end{figure}

\begin{figure}
\epsscale{1.14}
\plotone{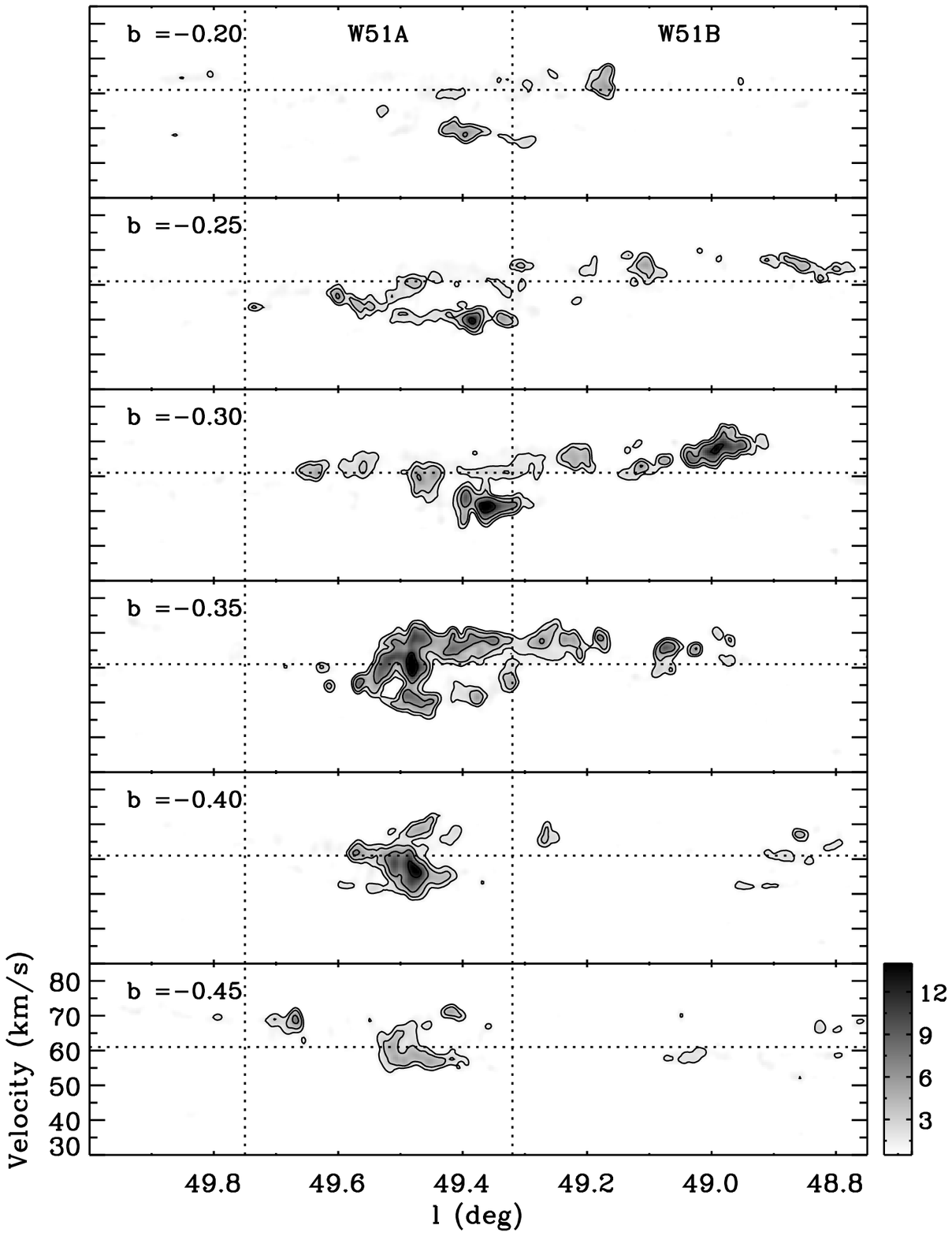}

\caption{The longitude-velocity map of \tco\ \jt\ along the latitude from
$b=-0\fdg20$ to $b=-0\fdg45$ by 0\fdg05 step. Contour levels are 1.5,
3, 6, and 12 K in $T_{MB}$. }

\label{lvmap_13co}
\end{figure}

\begin{figure}
\epsscale{1.14}
\plotone{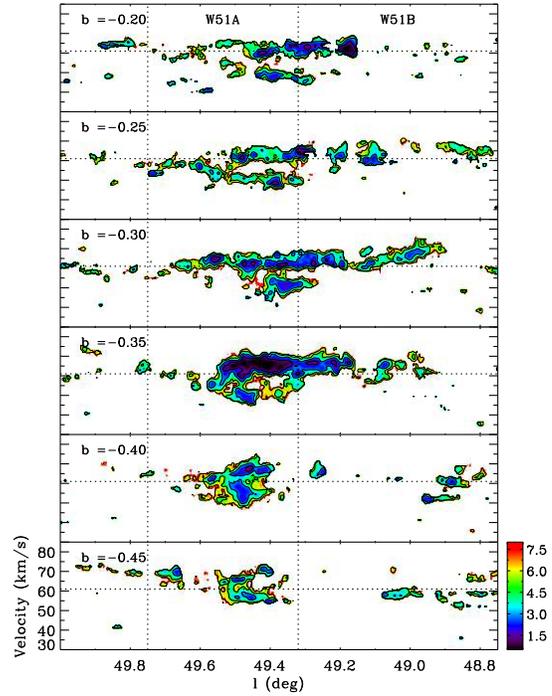}

\caption{The longitude-velocity map of $^{12/13}R_{2-1}$ along the
latitude from $b=-0\fdg20$ to $b=-0\fdg45$ by 0\fdg05 step. Contour
levels are 1.5, 3.0, 5.0, and 7.0.}

\label{lvmap_r1213}
\end{figure}

Comparing the velocities of our CO data to the velocities of RRL,
H$_2$CO and HI absorption \citep{Arnal85, Bieging75,Downes80, Koo97a,
Mufson79,Pankonin79} discussed in previous sections, we deduce the
spatial distribution of the W51 complex. The spatial distribution of
the \ion{H}{2} regions and YSOs around the central active star-forming
region is shown in Figure \ref{w51_small}. Many small clusters of 5-10
YSOs are apparent. Five of them are marked by large circles which happen
to be close to PV cuts. Figures \ref{lvmap_12co}--\ref{lvmap_r1213} show
the velocity structure of \co\ \jt, \tco\ \jt, and $^{12/13}R_{2-1}$
from $b=-0\fdg45$ to $b=-0\fdg20$ by a $0\fdg05$ step. We compare
the spatial distribution of the \ion{H}{2} regions and the YSO
clusters to the latitude-velocity ($l,v$) maps of \co\ \jt, \tco\
\jt, and their ratio ($^{12/13}R_{2-1}$) to determine the relative
positions of the molecular clouds associated with the \ion{H}{2}
and YSO cluster regions. The brightest \ion{H}{2} region, G49.5-0.4
extends from $b=-0\fdg45$ to $b=-0\fdg30$. The mean RRL velocity of
59 \kms\ agrees with the strong \co\ and \tco\ \jt\ intensity. Most
of the G49.4-0.3 sources are located on the $b=-0\fdg30$ line; their
mean RRL velocity is 52 \kms. As seen on the section of $b=-0\fdg30$
in Figures \ref{lvmap_12co}--\ref{lvmap_r1213}, two strong components
of molecular gas are located at 52 \kms\ and 68 \kms. Those are
associated with G49.4-0.3 and G49.0-0.3. The HV molecular components
over 70 \kms\ in W51A are not evident on the $b=-0\fdg30$ line. In
Figures \ref{lvmap_12co}--\ref{lvmap_r1213}, the dense molecular cloud
around ($l,v$) = (49\fdg40, 50 \kms) in $b=-0\fdg25$ is associated with
component of G49.4-0.3 and one around ($l,v$) = (49\fdg20, 61 \kms) in
$b=-0\fdg20$ is connected with G49.17-0.21. The radio continuum sources
and the associated molecular cloud in W51A are behind the HV streaming gas
considering H$_2$CO absorption line velocities. The HV streaming gas in
front of G49.5-0.4 is much brighter in CO and more extended than that of
G49.4-0.3. Spatially, G49.4-0.3 at 52 \kms\ is located behind G49.5-0.4
at 58 \kms. Most of the bright molecular clouds in W51B are within the
velocity range of the HV stream. Toward G49.0-0.3 and G48.9-0.3, no other
molecular clouds are observed in the lower velocity range. The sources
of W51B within the HV streaming gas are located nearest to the Sun.

\begin{figure}
\plotone{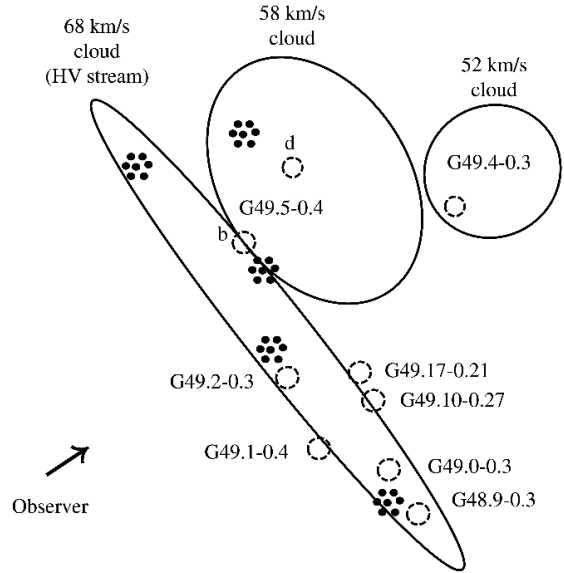}

\caption{Schematic diagram of the W51 \ion{H}{2} regions, YSO clusters,
and molecular clouds (compare with the diagram in \cite{Goudis82}). Dashed
circles and filled dots represent the \ion{H}{2} regions and YSO clusters,
respectively.}

\label{w51_schematic}
\end{figure}

YSOs showing clustering are embedded in their parent molecular
cloud. We find the positions of molecular clouds associated
with YSOs by comparing Figure \ref{w51_small} with Figures
\ref{lvmap_12co}--\ref{lvmap_r1213}. First, many YSOs are concentrated
around ($l,b = 49\fdg68, -0\fdg45$) in Figure \ref{w51_small}. A molecular
cloud is distributed around ($l,v$) = (49\fdg68, 70 \kms) in the ($l,v$)
map on $b=-0\fdg45$ (Figures \ref{lvmap_12co}--\ref{lvmap_r1213}). The
second YSO cluster around ($l,b =49\fdg62,-0\fdg25 $) is associated
with a molecular cloud around ($l,v$) = (49\fdg62, 58 \kms) on $b=-0\fdg25$
in Figures \ref{lvmap_12co}--\ref{lvmap_r1213}. The third YSO cluster
is located along the $b=-0\fdg35$ line between G49.5-0.4 and G49.4-0.3
(Figure \ref{w51_small}). In Figure \ref{lvmap_r1213}, the molecular
cloud around ($l,v$) = (49\fdg42, 65 \kms) at $b=-0\fdg35$ shows a
small ratio of $^{12/13}R_{2-1}$ where the optically thick region
coincides with the third YSO cluster. The fourth YSO cluster is near
G49.2-0.3 at ($l,v$) = (49\fdg28, 65 \kms) on $b=-0\fdg35$ in Figures
\ref{lvmap_12co}--\ref{lvmap_r1213}. The fifth YSO cluster is located
between G49.0-0.3 and G48.9-0.3 in Figure \ref{w51_small} and coincides
with the molecular cloud at ($l,v$) = (48\fdg95, 68 \kms) and $b=-0\fdg30$
in Figures \ref{lvmap_12co}--\ref{lvmap_r1213}. A schematic diagram
showing the distribution of molecular clouds, the \ion{H}{2} regions,
and the YSO clusters is shown in Figure \ref{w51_schematic}.

\subsection{Triggered Star Formation}
\label{sec:Triggered Star Formation}

In this section, we discuss triggered star formation in W51. Triggered
star formation was reviewed as three distinct mechanisms by
\cite{Elemegreen98}: (1) direct compression of clouds by supernovae,
stellar winds, and \ion{H}{2} regions \citep{Klein85}, (2) collect and
collapse \citep{Elmegreen77}, and (3) cloud collisions \citep{Loren76}.

The arc structure around G49.0-0.3 and G48.9-0.3 is likely to be
an example of star formation triggered by the collect and collapse
process. In a PV diagram through the two \ion{H}{2} regions (Figure
\ref{W51B_pv}), the molecular cloud comprises a redshifted arc structure
with $V_{LSR}$ = 60 to 73 \kms \citep{Moon98}. Two molecular clumps
in this arc structure coexist with two \ion{H}{2} regions. Another
molecular clump is located at higher velocity than the locations of the
two compact \ion{H}{2} regions. Newly forming YSOs are associated with
the dense molecular clumps (Figure \ref{G49.0_48.9-0.3_chanmap}). The
shell structure near the W51A region (Figure \ref{W51_rgb_co}) also shows
evidence of star formation triggered by the expansion of an \ion{H}{2}
region, as discussed in an earlier paper \citep{Kang09a}. Dense molecular
material has been collected along the shell detected in {\it Spitzer}
IRAC images. YSOs are identified toward the densest molecular condensation
along the edge of the IRAC shell.

\cite{Scoville86} suggested that most high-mass OB stars are formed
by cloud-cloud collisions based on the observational evidence for
many GMCs associated with \ion{H}{2} regions. The studies of star
formation triggered by cloud-cloud collisions have been reported in other
regions, e.g., NGC 1333 \citep{Loren76}, DR 21-W75 \citep{Dickel78}, W49N
\citep{Serabyn93}, and Sagittarius B2 \citep{Hasegawa94, Mehringer93}. The
possibility of triggered star formation by cloud-cloud collisions in W51
has already been suggested \citep{Pankonin79, Arnal85, Carpenter98}.

Cloud-cloud collisions may compress the interface region between two
colliding clouds and result in the initiation of star formation. The
molecular clouds at this interface are heated, but the molecular clouds on
the trailing side remain cold. As a result, self-absorption of CO lines
is observed toward the interface of cloud-cloud collisions. Models for
collisions between two different clouds have shown that a large cloud can
be disrupted by the bow shock induced by the collision with a small cloud,
but the small cloud is compressed by the shock, and then the post-shock
gas of the small cloud can form a new star \citep{Habe92}.

In Figures \ref{lvmap_12co}--\ref{lvmap_r1213}, the velocity of 61 \kms\
divides the HV stream from the W51 main components. The HV components
are related to the streaming motions in the Sagittarius spiral arm. Their
elongated structure may result from the associated spiral-wave shock. We
agree with earlier suggestions that active massive star formation in
the W51 \ion{H}{2} region complex is due to a collision between the
molecular clouds in the HV stream and the molecular cloud below 61 \kms\
\citep{Pankonin79, Arnal85, Carpenter98}. In the regions G49.17-0.21
and G49.5-0.4, the self-absorption of \co\ \jt\ occurs around 65 \kms,
coincident with the dark shadow region in IRAC images. In addition,
several YSOs are detected inside the low ratio regions ($^{12/13}R_{2-1}
< 1.5$). Dark filamentary structures in IR bands (IRDCs) are believed
to indicate the earliest stages of massive star-forming regions
\citep{Beuther07}. The IRDCs associated with G49.17-0.21 and G49.5-0.4
have kinematics revealed by the CO maps that appear to be consistent
with the cloud collision model for massive star formation.

Star formation in the W51 complex appears to have occurred in a
long and large filament along with the \ion{H}{2} regions (Figure
\ref{12co_lb}). We could not find clear evidence for a time sequence
in this formation process from our data. YSOs along the filament,
however, are younger and more massive than those in the surrounding
region \citep[see][Figures 6 and 7]{Kang09b}.

\subsection{Star Formation Efficiency and Rate}
\label{sec:SFE}

Estimating the masses of stars and their parent molecular clouds is
important in understanding how molecular clouds evolve and form stars.
The star formation efficiency (SFE) is defined as $SFE = M_{*} / (M_{gas}+
M_{*})$, where $M_{gas}$ is the mass of cloud and $M_{*}$ is the mass of
stars in the current generation. \cite{Williams97} have found that 10\%
of the mass of a GMC is converted to stars. \cite{Koo99} reported SFE
values of 7\% and 15\% in the northwest and southeast components of W51B.

We can compare the gas mass with the mass currently identified as YSOs. We
predict that the extended total YSO mass of W51 (W51A + W51B) is $2.1
\times 10^{3} M_\odot$ using the mass functions from \cite{Kang09b} by
extrapolating the fitted power laws down to 1 $M_\odot$. Total masses
of YSOs in W51A and W51B are $0.9 \times 10^{3} M_\odot$ and $1.2 \times
10^{3} M_\odot$, respectively. As discussed in Section \ref{sec:physical
properties} we used the LVG formalism for radiative transfer of CO lines
in molecular clouds to estimate the gas mass. Total gas masses of each
region are listed in Table \ref{CO.tbl.masses}. The current YSO mass
fractions in W51A and W51B are 0.7\% and 1.1 \%, respectively.

These estimates are for the mass of YSOs {\it currently} detected in
the selected region by extrapolating in mass over our mass function to
undetected YSOs. The overall star formation efficiency is the ratio of
the mass of stars formed over the cloud lifetime ($T_{cloud}$) to the
initial cloud mass, or approximately, to the current cloud mass plus the
mass of all stars formed up to now. We find that only about 1\% of the
cloud mass is currently in YSOs. We assume the mean timescale for a YSO
in the selected mass range to be detected by IRAC ($T_{YSO}$) is $\sim
3\times10^4$ yr, which is the statistical lifetime of high-mass protostars
in the Cygnus X molecular cloud complex \citep{Motte07}.  If the final
star formation efficiency is more like 10\% \citep{Williams97,Koo99}
the star formation process must continue for about $10 \times T_{YSO}$
at the present rate to convert 10\% of the cloud to stars. So the current
fraction of YSOs, together with an assumed YSO timescale implies the
overall time during which the cloud is actively forming stars. If the YSO
timescale for high mass stars is so short, $3\times10^4$ yr, for a typical
cloud lifetime of $3\times10^6$ yr and star formation occurred at the
presently observed rate, the W51 clouds would convert 10\% of their mass
to stars in only 1/10th of the cloud lifetime. We are probably observing
these clouds at an exceptionally high star formation rate, relative to
the average over the cloud lifetime, which typically does not exceed 5 --
10 Myrs \citep{Leisawitz89}. That is consistent with the remarkably high
content of massive YSOs that we found in \cite{Kang09b}. This ``burst
of star formation'' may be a consequence of the passage of the W51 GMC
through the spiral arm.

\section{Summary}
\label{co:summary}

We observed the molecular clouds associated with the active star-forming
complex W51 in the \co\ and \tco\ \jt\ lines with the HHT. We summarize
our results, combining the new CO data of high sensitivity and resolution,
together with {\it Spitzer} 3.6 - 24 \micron\ data:

1. CO emission in the 1\fdg25 $\times$ 1\fdg00 area divides into three
velocity-components, $30-55$, $56-65$, and $66-85$ \kms. G49.4-0.3 and
some clouds in the northwestern part of the map are associated with the
lower velocity ($30-55$ \kms) component. The main molecular clouds in
the velocity range of 56 to 65 \kms\ are distributed widely throughout
the whole region. The higher velocity component ($> 66$ \kms) shows an
elongated filamentary structure from southeast to northwest.

2. The average ratios of $^{13}R_{2-1/1-0}$ and $^{13/12}R_{2-1}$ of
all \ion{H}{2} regions are greater than the values outside the active
star-forming regions, which implies that the molecular gas directly
associated with the \ion{H}{2} regions is dense and highly excited.

3. We compare our CO maps with {\it Spitzer} data. Strong PAH emission
near \ion{H}{2} regions seen in IRAC bands coincides with bright CO
emission, suggesting that those molecular clouds are associated with
the \ion{H}{2} regions. Many YSOs are detected in the dark parts near
the \ion{H}{2} regions on IRAC images.

4. We argue that triggered star formation resulted from \ion{H}{2} region
expansion into clouds in the shell structure near W51A and in the region
around G49.0-0.3 and G48.9-0.3. We also present evidence of star formation
triggered by cloud-cloud collisions in G49.5-0.4 and G49.17-0.21.

5. We estimate the total gas masses of the various interesting regions
using an LVG analysis of \co\ and \tco\ \jt\ data. By comparing the
total gas mass to total mass of YSOs in W51A and W51B, we find current
YSO formation efficiencies of 0.7\% and 1.1\%, respectively, within a
timescale of $3 \times 10^4$ yr during which massive stars ($>5$\msun)
would be detected as YSOs. The {\it current} rate of star formation in
the W51 GMC should then be an order of magnitude higher than the rate
averaged over an assumed cloud lifetime of $3 \times 10^6$ yr, in order
to convert 10\% of the cloud mass to stars.

\acknowledgments
We thank Bon-Chul Koo for providing 21 cm radio continuum data. This
research was supported in part by NSF grant AST-0708131 to the University
of Arizona, and by the Korea Research Foundation Grant funded by the
Korean Government (MOEHRD: KRF-2007-612C00050). MC was supported by the
International Research \& Development Program of the National Research
Foundation of Korea (NRF) funded by the Ministry of Education, Science and
Technology (MEST) of Korea (Grant number: K20901001400-09B1300-03210, FY
2009). This publication makes use of molecular line data from the Boston
University-FCRAO Galactic Ring Survey (GRS), which is a joint project
of Boston University and Five College Radio Astronomy Observatory,
funded by the National Science Foundation under grants AST-9800334,
AST-0098562, \& AST-0100793.


\begin{thebibliography}{}

 \bibitem[Arnal \& Goss(1985)]{Arnal85} Arnal, E.~M., \& Goss, W.~M.\ 1985, \aap, 145, 369  
 \bibitem[Benjamin et al.(2003)]{Benjamin03} Benjamin, R.~A., et al.\ 2003, \pasp, 115, 953  
 \bibitem[Beuther et al.(2007)]{Beuther07} Beuther, H., Churchwell, E.~B., McKee, C.~F., \& Tan, J.~C.\ 2007, in Protostars and Planets V, ed. B. Reipurth, D. Jewitt, \& K. Keil (Tucson, AZ: Univ. Arizona Press), 165  
 \bibitem[Bieging(1975)]{Bieging75} Bieging, J.\ 1975, \ in ``H II regions and related topics'', eds. T.L. Wilson \& D. Downes, Lecture Notes in Physics, Vol. 42 (Berlin: Springer), p. 443
 \bibitem[Bieging et al.(2009)]{Bieging09} Bieging, J.~H., Peters, W.~L., Vila Vilaro, B., Schlottman, K., \& Kulesa, C.\ 2009, \aj, 138, 975 
 \bibitem[Black \& van Dishoeck(1987)]{Black87} Black, J.~H., \& van Dishoeck, E.~F.\ 1987, \apj, 322, 412 
 \bibitem[Brand \& Blitz(1993)]{Brand93} Brand, J., \& Blitz, L.\ 1993, \aap, 275, 67 
 \bibitem[Bohlin et al.(1978)]{Bohlin78} Bohlin, R.~C., Savage, B.~D., \& Drake, J.~F.\ 1978, \apj, 224, 132  
 \bibitem[Burton(1970)]{Burton70} Burton, W.~B.\ 1970, \aaps, 2, 291 
 \bibitem[Cardelli et al.(1996)]{Cardelli96} Cardelli, J.~A., Meyer, D.~M., Jura, M., \& Savage, B.~D.\ 1996, \apj, 467, 334 
 \bibitem[Carey et al.(2005)]{Carey05} Carey, S.~J., et al.\ 2005, Bulletin of the American Astronomical Society, 37, 1252 
 \bibitem[Carpenter \& Sanders(1998)]{Carpenter98} Carpenter, J.~M., \& Sanders, D.~B.\ 1998, \aj, 116, 1856 
 \bibitem[Clark \& Porter(2004)]{Clark04} Clark, J.~S., \& Porter, J.~M.\ 2004, \aap, 427, 839 
 \bibitem[Crampton et al.(1978)]{Crampton78} Crampton, D., Georgelin, Y.~M., \& Georgelin, Y.~P.\ 1978, \aap, 66, 1  
 \bibitem[Cyganowski et al.(2008)]{Cyganowski08} Cyganowski, C.~J., et al.\ 2008, \aj, 136, 2391 
 \bibitem[Davis et al.(2007)]{Davis07} Davis, C.~J., Kumar, M.~S.~N., Sandell, G., Froebrich, D., Smith, M.~D., \& Currie, M.~J.\ 2007, \mnras, 374, 29  
 \bibitem[Dickel et al.(1978)]{Dickel78} Dickel, J.~R., Dickel, H.~R., \& Wilson, W.~J.\ 1978, \apj, 223, 840  
 \bibitem[Downes et al.(1980)]{Downes80} Downes, D., Wilson, T.~L., Bieging, J., \& Wink, J.\ 1980, \aaps, 40, 379  
 \bibitem[Elmegreen \& Lada(1977)]{Elmegreen77} Elmegreen, B.~G., \& Lada, C.~J.\ 1977, \apj, 214, 725 
 \bibitem[Elmegreen(1998)]{Elemegreen98} Elmegreen, B.~G.\ 1998, Origins, ed. C. E. Woodward, J. M. Shull, \& H. A. Thronson, Jr. (San Francisco, CA: ASP), 150
 \bibitem[Fazio et al.(2004)]{Fazio04} Fazio, G.~G., et al.\ 2004, \apjs, 154, 10   
 \bibitem[Figuer{\^e}do et al.(2008)]{Figueredo08} Figuer{\^e}do, E., Blum, R.~D., Damineli, A., Conti, P.~S., \& Barbosa, C.~L.\ 2008, \aj, 136, 221 
 \bibitem[Fomalont \& Weliachew(1973)]{Fomalont73} Fomalont, E.~B., \& Weliachew, L.\ 1973, \apj, 181, 781  
 \bibitem[Genzel \& Downes(1977)]{Genzel77} Genzel, R., \& Downes, D.\ 1977, \aaps, 30, 145  
 \bibitem[Genzel et al.(1981)]{Genzel81} Genzel, R., et al.\ 1981, \apj, 247, 1039 
 \bibitem[Goss \& Shaver(1970)]{Goss70} Goss, W.~M., \& Shaver, P.~A.\ 1970, Australian Journal of Physics Astrophysical Supplement, 14, 1
 \bibitem[Goudis \& Hippelein(1982)]{Goudis82} Goudis, C., \& Hippelein, H.\ 1982, \aap, 105, 329  
 \bibitem[Gritschneder et al.(2009)]{Gritschneder09} Gritschneder, M., Naab, T., Walch, S., Burkert, A., \& Heitsch, F.\ 2009, \apjl, 694, L26 
 \bibitem[Habe \& Ohta(1992)]{Habe92} Habe, A., \& Ohta, K.\ 1992, \pasj, 44, 203  
 \bibitem[Hasegawa et al.(1994)]{Hasegawa94} Hasegawa, T., Sato, F., Whiteoak, J.~B., \& Miyawaki, R.\ 1994, \apjl, 429, L77 
 \bibitem[Imai et al.(2002)]{Imai02} Imai, H., Watanabe, T., Omodaka, T., Nishio, M., Kameya, O., Miyaji, T., \& Nakajima, J.\ 2002, \pasj, 54, 741 
 \bibitem[Israel(1978)]{Israel78} Israel, F.~P.\ 1978, \aap, 70, 769  
 \bibitem[Jackson et al.(2006)]{Jackson06} Jackson, J.~M., et al.\ 2006, \apjs, 163, 145 
 \bibitem[Kang et al.(2009a)]{Kang09a} Kang, M., Bieging, J.~H., Kulesa, C.~A., \& Lee, Y.\ 2009a, \apj, 701, 454  
 \bibitem[Kang et al.(2009b)]{Kang09b} Kang, M., Bieging, J.~H., Povich, M.~S., \& Lee, Y.\ 2009b, \apj, 706, 83 
 \bibitem[Klein et al.(1985)]{Klein85} Klein, R.~I., Whitaker, R.~W., \& Sandford, M.~T., II 1985, Protostars and Planets II, 340 
 \bibitem[Koo \& Moon(1997)]{Koo97} Koo, B.-C., \& Moon, D.-S.\ 1997, \apj, 475, 194  
 \bibitem[Koo(1997)]{Koo97a} Koo, B.-C.\ 1997, \apjs, 108, 489  
 \bibitem[Koo(1999)]{Koo99} Koo, B.-C.\ 1999, \apj, 518, 760  
 \bibitem[Kulesa et al.(2005)]{Kulesa05} Kulesa, C.~A., Hungerford, A.~L., Walker, C.~K., Zhang, X., \& Lane, A.~P.\ 2005, \apj, 625, 194  
 \bibitem[Kumar et al.(2004)]{Kumar04} Kumar, M.~S.~N., Kamath, U.~S., \& Davis, C.~J.\ 2004, \mnras, 353, 1025 
 \bibitem[Kundu \& Velusamy(1967)]{Kundu67} Kundu, M.~R., \& Velusamy, T.\ 1967, Annales d'Astrophysique, 30, 59  
 \bibitem[Kutner \& Ulich(1981)]{Kutner81} Kutner, M.~L., \& Ulich, B.~L.\ 1981, \apj, 250, 341 
 \bibitem[Leisawitz et al.(1989)]{Leisawitz89} Leisawitz, D., Bash, F.~N., \& Thaddeus, P.\ 1989, \apjs, 70, 731  
 \bibitem[Loren(1976)]{Loren76} Loren, R.~B.\ 1976, \apj, 209, 466  
 \bibitem[Mehringer et al.(1993)]{Mehringer93} Mehringer, D.~M., Palmer, P., Goss, W.~M., \& Yusef-Zadeh, F.\ 1993, \apj, 412, 684  
 \bibitem[Mehringer(1994)]{Mehringer94} Mehringer, D.~M.\ 1994, \apjs, 91, 713  
 \bibitem[Moon \& Park(1998)]{Moon98} Moon, D.-S., \& Park, Y.-S.\ 1998, \mnras, 296, 863  
 \bibitem[Motte et al.(2007)]{Motte07} Motte, F., Bontemps, S., Schilke, P., Schneider, N., Menten, K.~M., \& Brogui{\`e}re, D.\ 2007, \aap, 476, 1243  
 \bibitem[Mufson \& Liszt(1979)]{Mufson79} Mufson, S.~L., \& Liszt, H.~S.\ 1979, \apj, 232, 451  
 \bibitem[Pankonin et al.(1979)]{Pankonin79} Pankonin, V., Payne, H.~E., \& Terzian, Y.\ 1979, \aap, 75, 365  
 \bibitem[Penzias et al.(1971)]{Penzias71} Penzias, A.~A., Solomon, P.~M., Wilson, R.~W., \& Jefferts, K.~B.\ 1971, \apjl, 168, L53  
 \bibitem[Povich et al.(2009)]{Povich09} Povich, M.~S., et al.\ 2009, \apj, 696, 1278 
 \bibitem[Rieke et al.(2004)]{Rieke04} Rieke, G.~H., et al.\ 2004, \apjs, 154, 25 
 \bibitem[Robitaille et al.(2007)]{Robitaille07} Robitaille, T.~P., Whitney, B.~A., Indebetouw, R., \& Wood, K.\ 2007, \apjs, 169, 328 
 \bibitem[Russeil(2003)]{Russeil03} Russeil, D.\ 2003, \aap, 397, 133 
 \bibitem[Sakamoto et al.(1997)]{Sakamoto97} Sakamoto, S., Hasegawa, T., Handa, T., Hayashi, M., \& Oka, T.\ 1997, \apj, 486, 276  
 \bibitem[Sanders et al.(1986)]{Sanders86} Sanders, D.~B., Clemens, D.~P., Scoville, N.~Z., \& Solomon, P.~M.\ 1986, \apjs, 60, 1  
 \bibitem[Sault et al.(1995)]{Sault95} Sault, R.~J.,Teuben, P.~J., \& Wright, M.~C.~H.\ 1995, in ASP Conf. Ser. 77, Astronomical Data Analysis Software and Systems IV, ed. R. A. Shaw, H. E. Payne, \& J. J. E. Hayes (San Francisco, CA: ASP), 433 
 \bibitem[Schneps et al.(1981)]{Schneps81} Schneps, M.~H., Lane, A.~P., Downes, D., Moran, J.~M., Genzel, R., \& Reid, M.~J.\ 1981, \apj, 249, 124 
 \bibitem[Scoville et al.(1986)]{Scoville86} Scoville, N.~Z., Sanders, D.~B., \& Clemens, D.~P.\ 1986, \apjl, 310, L77  
 \bibitem[Serabyn et al.(1993)]{Serabyn93} Serabyn, E., Guesten, R., \& Schulz, A.\ 1993, \apj, 413, 571  
 \bibitem[Shepherd et al.(2007)]{Shepherd07} Shepherd, D.~S., et al.\ 2007, \apj, 669, 464  
 \bibitem[Skrutskie et al.(2006)]{Skrutskie06} Skrutskie, M.~F., et al.\ 2006, \aj, 131, 1163 
 \bibitem[Smith et al.(2006)]{Smith06} Smith, H.~A., Hora, J.~L., Marengo, M., \& Pipher, J.~L.\ 2006, \apj, 645, 1264  
 \bibitem[Snyder \& Buhl(1971)]{Snyder71} Snyder, L.~E., \& Buhl, D.\ 1971, \apjl, 163, L47  
 \bibitem[Subrahmanyan \& Goss(1995)]{Subrahmanyan95} Subrahmanyan, R., \& Goss, W.~M.\ 1995, \mnras, 275, 755  
 \bibitem[van Dishoeck \& Black(1988)]{vanDishoeck88} van Dishoeck, E.~F., \& Black, J.~H.\ 1988, \apj, 334, 771  
 \bibitem[van Gorkom et al.(1980)]{vanGorkom80} van Gorkom, J.~H., Goss, W.~M., Shaver, P.~A., Schwarz, U.~J., \& Harten, R.~H.\ 1980, \aap, 89, 150  
 \bibitem[Walborn et al.(2002)]{Walborn02} Walborn, N.~R., Ma{\'{\i}}z-Apell{\'a}niz, J., \& Barb{\'a}, R.~H.\ 2002, \aj, 124, 1601 
 \bibitem[Williams \& McKee(1997)]{Williams97} Williams, J.~P., \& McKee, C.~F.\ 1997, \apj, 476, 166  
 \bibitem[Wilson et al.(1970)]{Wilson70} Wilson, T.~L.,Mezger, P.~G., Gardner, F.~F., \& Milne, D.~K.\ 1970, \aplett, 5, 99  
 \bibitem[Wynn-Williams et al.(1974)]{Wynn-Williams74} Wynn-Williams, C.~G., Becklin, E.~E., \& Neugebauer, G.\ 1974, \apj, 187, 473  
 \bibitem[Xu et al.(2009)]{Xu09} Xu, Y., Reid, M.~J., Menten, K.~M., Brunthaler, A., Zheng, X.~W., \& Moscadelli, L.\ 2009, \apj, 693, 413 
\end{thebibliography}
\end{document}